\documentclass[sigconf]{acmart}
\usepackage{subfigure}
\usepackage[norelsize,ruled,vlined,linesnumbered]{algorithm2e}
\usepackage{algorithmic}
\usepackage{url}
\usepackage{color}

\AtBeginDocument{%
  \providecommand\BibTeX{{%
    \normalfont B\kern-0.5em{\scshape i\kern-0.25em b}\kern-0.8em\TeX}}}

\newcommand{\wsq}{\hspace{\fill}$\square$}
\newcommand{\vs}{\vspace{1.5mm}}
\newcommand{\argmax}{\mathop{\rm arg\, max}\limits}
\newcommand{\argmin}{\mathop{\rm arg\, min}\limits}

\newtheoremstyle{mystyle}
    {2.0mm}
    {2.0mm}
    {\it}
    {0mm}
    {\scshape}
    {.}
    { }
    {}
\theoremstyle{mystyle}
\newtheorem{defi}{Definition}
\newtheorem{exa}{Example}
\newtheorem{lemm}{Lemma}
\newtheorem{theo}{Theorem}
\newtheorem{coro}{Corollary}
\newtheorem{assumption}{Assumption}
\newtheorem{remark}{Remark}

\setcopyright{acmcopyright}
\copyrightyear{}
\acmYear{}
\acmDOI{}
\acmConference[]{}{}{}
\acmBooktitle{}
\acmPrice{}
\acmISBN{}

\setcopyright{none}
\renewcommand\footnotetextcopyrightpermission[1]{} 
\begin{document}
\fancyhead{}

\title{Fast Density-Peaks Clustering: Multicore-based Parallelization Approach}

\author{Daichi Amagata}
\affiliation{%
\institution{Osaka University, PRESTO}
\country{Japan}}
\email{amagata.daichi@ist.osaka-u.ac.jp}

\author{Takahiro Hara}
\affiliation{%
\institution{Osaka University}
\country{Japan}}
\email{hara@ist.osaka-u.ac.jp}

\begin{abstract}
Clustering multi-dimensional points is a fundamental task in many fields, and density-based clustering supports many applications as it can discover clusters of arbitrary shapes.
This paper addresses the problem of Density-Peaks Clustering (DPC), a recently proposed density-based clustering framework.
Although DPC already has many applications, its straightforward implementation incurs a quadratic time computation to the number of points in a given dataset, thereby does not scale to large datasets.

To enable DPC on large datasets, we propose efficient algorithms for DPC.
Specifically, we propose an exact algorithm, Ex-DPC, and two approximation algorithms, Approx-DPC and S-Approx-DPC.
Under a reasonable assumption about a DPC parameter, our algorithms are sub-quadratic, i.e., break the quadratic barrier.
Besides, Approx-DPC does not require any additional parameters and can return the same cluster centers as those of Ex-DPC, rendering an accurate clustering result.
S-Approx-DPC requires an approximation parameter but can speed up its computational efficiency.
We further present that their efficiencies can be accelerated by leveraging multicore processing.
We conduct extensive experiments using synthetic and real datasets, and our experimental results demonstrate that our algorithms are efficient, scalable, and accurate.
\end{abstract}

\maketitle

\section{Introduction}	\label{section_introduction}
Given a set $P$ of $n$ points in a $d$-dimensional space, clustering them aims at dividing $P$ into some subsets, i.e., clusters.
This multi-dimensional point clustering is a fundamental task for many data mining applications.
Density-based clustering particularly supports them well because it can discover clusters of arbitrary shapes.
This is an advantage over other clustering such as $k$-means \cite{gan2015dbscan, zhang2016efficient}.

\vs
\noindent
\underline{\textbf{DPC Framework.}}
Recently, as a novel density-based clustering framework, \textit{Density-Peaks Clustering} (DPC) has been proposed in \cite{rodriguez2014clustering}, and we focus on DPC in this paper.
Given a point set $P$, DPC computes, for each point $p_{i} \in P$,
\begin{itemize}
    \setlength{\leftskip}{-4.0mm}
    \setlength{\itemsep}{0.0mm}
    \item	\textit{local density} $\rho_{i}$: the number of points $p_{j}$ whose distances between $p_{i}$ and $p_{j}$ are less than $d_{cut}$, which is a user-specified cutoff parameter, and
    \item	\textit{dependent distance} $\delta_{i}$: the distance from $p_{i}$ to its nearest neighbor point in $P$ with higher local density than $\rho_{i}$ (i.e., \textit{dependent point}).
\end{itemize}
Then DPC identifies
\begin{itemize}
    \setlength{\leftskip}{-4.0mm}
    \setlength{\itemsep}{0.0mm}
    \item	\textit{noises}: points with less local density than $\rho_{min}$, and
    \item	\textit{cluster centers}: non-noise points, whose dependent distances are at least $\delta_{min}$ (each cluster center should have a comparatively long dependent distance, as its local              density is \textit{peak} at its area).
\end{itemize}
After that, the remaining points are assigned to the same cluster as their dependent points.

\begin{table*}[!t]
    \begin{center}
	\caption{The time complexity of each algorithm on a single thread for fixed dimensionality $d$ (under a reasonable assumption).
        Note that $n$ is the cardinality of a given dataset, $M$ is the number of compound LSHs, $b$ is the average bucket size of an LSH, and $\rho_{avg}$ is the average local density.}
        \label{table_time-complexity}
	\begin{tabular}{l||c|c|c} \hline
    	Algorithm							& Local density computation			& Dependent distance computation	& Total								\\ \hline \hline
            Scan								& $O(n^2)$							& $O(n^2)$							& $O(n^2)$							\\ \hline
            LSH-DDP \cite{zhang2016efficient}	& $O(M\sum b^2)$					& $O(M\sum b^2)$					& $O(M\sum b^2)$					\\ \hline
            CFSFDP-A \cite{bai2017fast}			& $O(n^2)$							& $\Omega(n^2)$						& $\Omega(n^2)$						\\ \hline
            Our algorithms in this paper		& $O(n(n^{1-1/d} + \rho_{avg}))$	& $O(n^{2-1/d})$					& $O(n(n^{1-1/d} + \rho_{avg}))$	\\ \hline
	\end{tabular}
    \end{center}
\end{table*}

\vs
\noindent
\underline{\textbf{Advantages of DPC.}}
In addition to the inherent advantages of density-based clustering, DPC has some advantages.
One of them is that, even if users (or applications) are not domain experts, they can intuitively select cluster centers and noises from a \textit{decision graph}, which visualizes $\langle \rho_{i},\delta_{i}\rangle$ into a 2-dimensional space.

\begin{figure}[!t]
    \begin{center}
	\subfigure[Dataset S2]{%
		\includegraphics[width=0.480\linewidth]{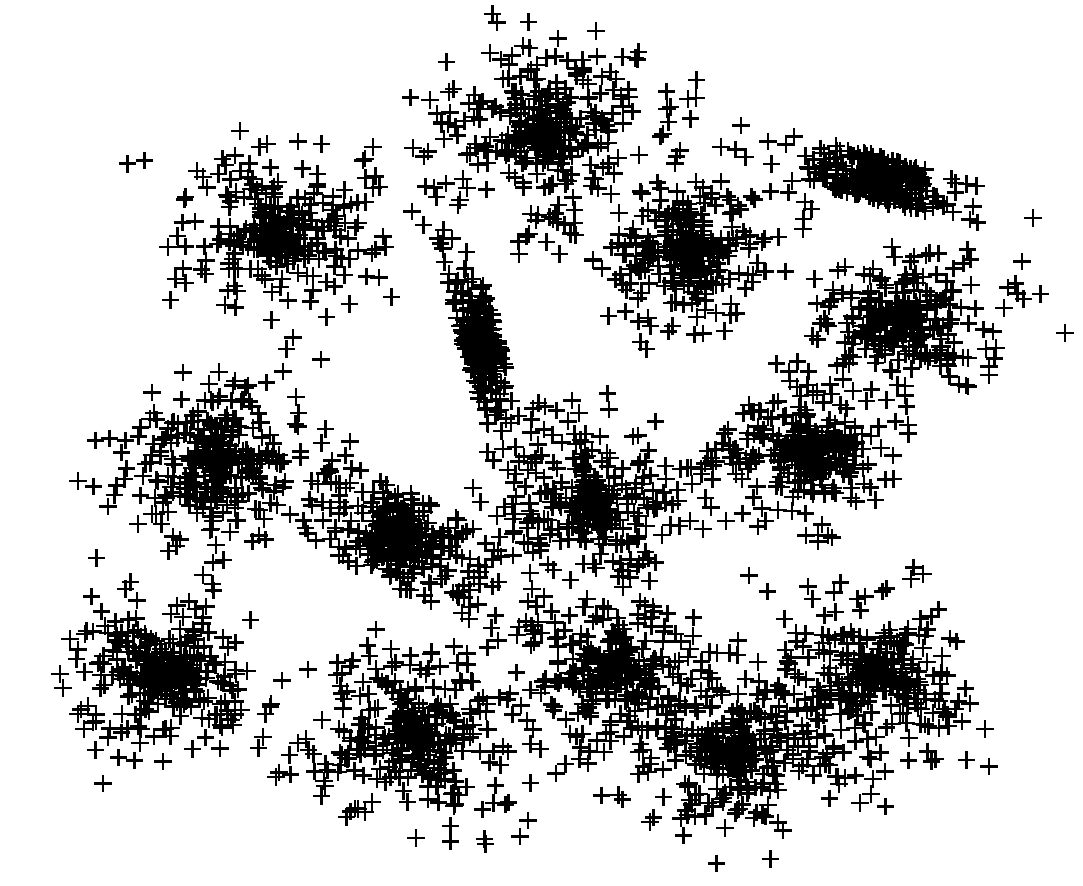}	\label{fig_s2}}
        \subfigure[Decision graph of S2]{%
		\includegraphics[width=0.480\linewidth]{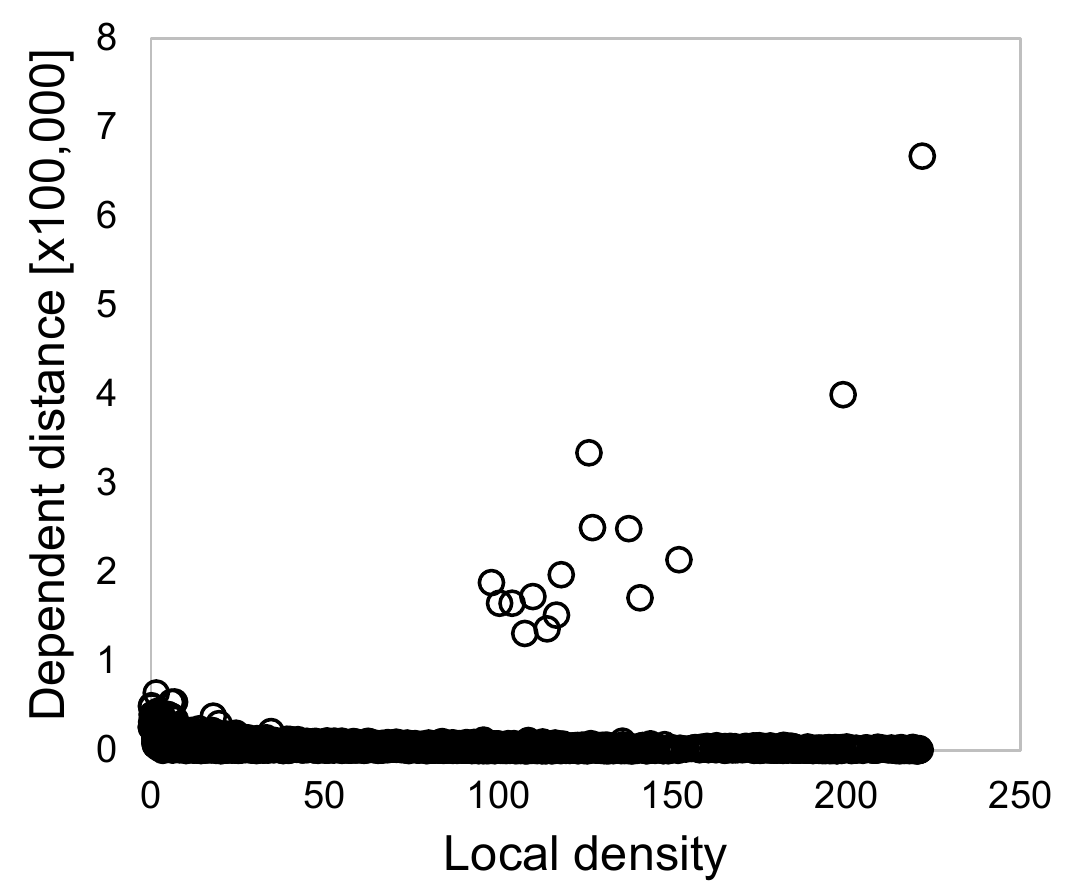}	\label{fig_decision-graph}}
        \caption{Illustration of S2 and its decision graph}	\label{figure_s2}
    \end{center}
\end{figure}

\begin{exa}[]	\label{example_decision-graph}
Figure \ref{figure_s2} illustrates (a) dataset S2 \cite{franti2018k} and (b) its decision graph.
S2 has 15 Gaussian clusters.
The decision graph depicts that 15 points have comparatively large dependent distances.
That is, it visually suggests that there are 15 clusters in the dataset and does a threshold ($\delta_{min}$) for obtaining their centers (recall that cluster centers tend to have comparatively long dependent distances).
Thanks to this observation, users can easily specify $\delta_{min}$ and $\rho_{min}$.
\end{exa}

\begin{figure}[!t]
    \begin{center}
	\subfigure[DPC]{%
		\includegraphics[width=0.480\linewidth]{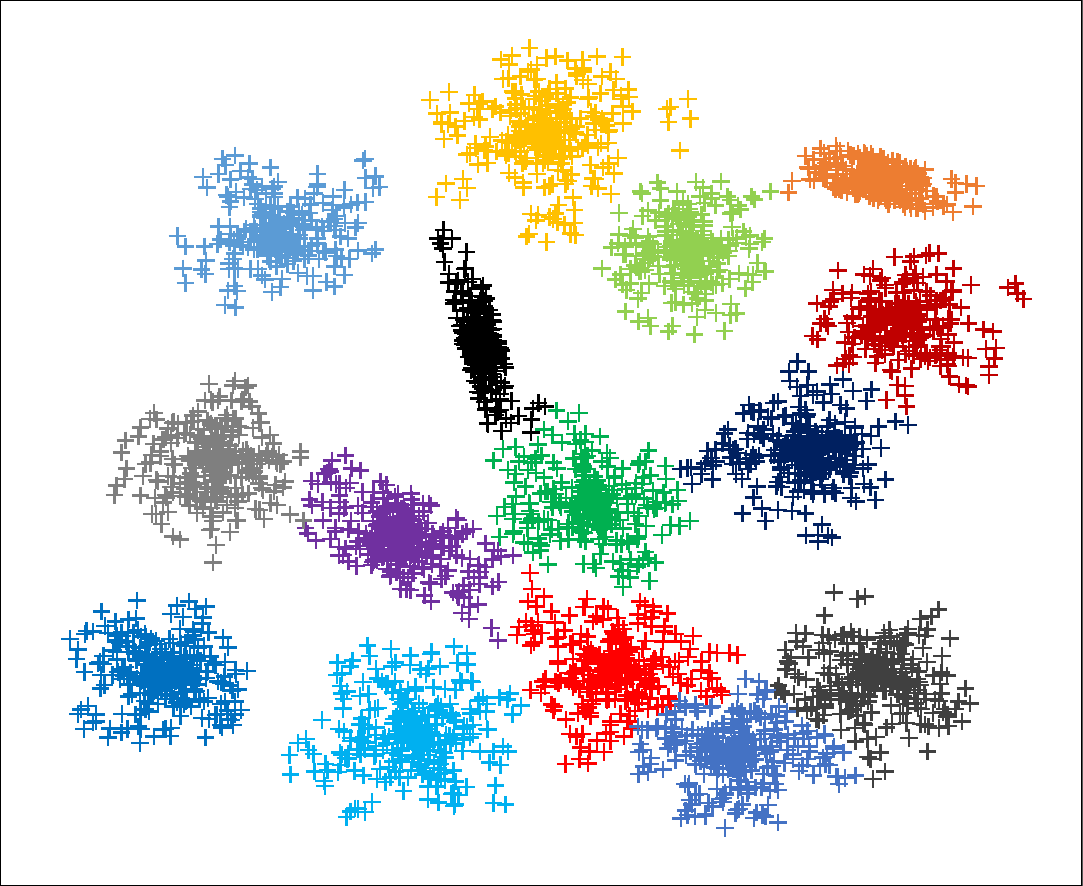}		\label{fig_s2-dpc}}
        \subfigure[DBSCAN]{%
		\includegraphics[width=0.480\linewidth]{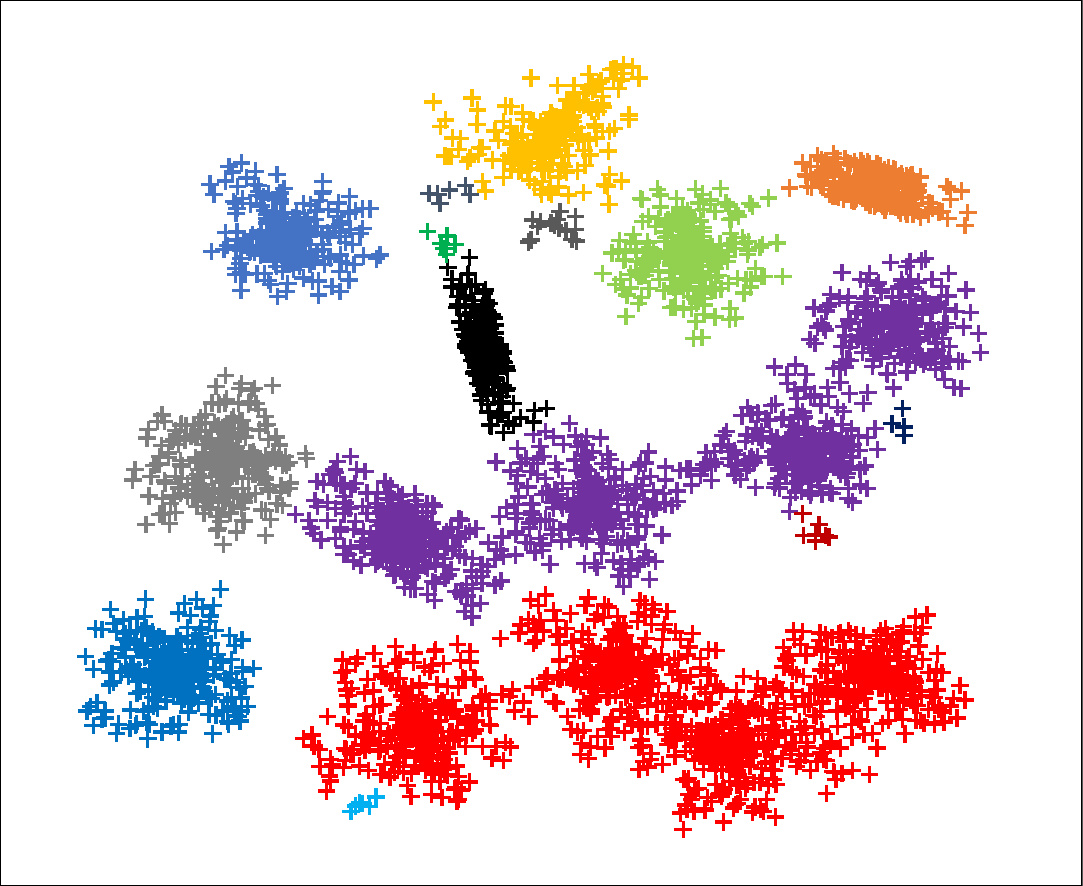}	\label{fig_s2-dbscan}}
        \caption{Difference of the clustering results (clustering quality comparison) between DPC and DBSCAN on S2}
        \label{figure_s2-cluster}
    \end{center}
\end{figure}

\noindent
Another advantage of DPC is that it can divide a dense space into some sub-spaces if the space has some density-peaks.
On the other hand, a famous density-based clustering DBSCAN \cite{ester1996density} (and its variants) cannot deal with this case well, for example:

\begin{exa}	\label{example_optics}
Figure \ref{figure_s2-cluster} illustrates the clustering results of DPC and DBSCAN on S2 (best viewed in color).
The parameters of DBSCAN are specified so that 15 clusters are obtained from OPTICS \cite{ankerst1999optics}.
In this figure, DBSCAN fails to obtain the correct clusters (some clusters are merged to their neighbor clusters), whereas DPC does not.
\end{exa}

\noindent
This example presents that DPC is robust to data distributions, and \cite{bai2017fast, zhang2016efficient} have also confirmed this observation (on other datasets).

\vs
\noindent
\underline{\textbf{Motivation.}}
Real-life applications often generate datasets with arbitrary shaped clusters that may not be clearly separated \cite{chen2019adaptive} (e.g., they may have points existing between close clusters), and DPC deals with them well even if they have such clusters.
Besides this, DPC supports easy selection of cluster centers and noises (see Example \ref{example_decision-graph}).
DPC therefore has been already employed in many fields, e.g., neuroscience \cite{mehmood2017clustering}, medical application \cite{ulanova2016clustering}, document summarization \cite{zhang2015clustering}, graphics \cite{hu2017co}, and computer vision \cite{wang2018semi}.
This fact observes the importance of DPC.

The above applications obviously require fast DPC algorithms to deal with a large number of points.
However, the computational efficiency of DPC has not been tackled much so far.

\vs
\noindent
\textit{Suffering from the quadratic computation time.}
A straightforward computation of the local density and dependent distance (point) for a point $p_{i} \in P$ is to scan the whole $P$, which incurs $O(n)$ time.
Therefore, computing the local density and dependent distance for all points in $P$ incurs $O(n^2)$ time.
Some existing works \cite{bai2017fast, zhang2016efficient} reduced the practical computation time, but they still suffer from the quadratic computation time.

\vs
\noindent
\textit{Non parallel-friendly.}
Leveraging parallel processing environments is a practical approach to scale well to large datasets.
Nowadays many CPU cores (or threads) are available in a single machine, so multicore-based parallel algorithms for many database operations have been considered \cite{faleiro2017latch, ren2016design}.
As clustering is an important database and data mining operation that needs to deal with large datasets, a parallel-friendly DPC algorithm is motivated.
Although there exist parallel DPC algorithms \cite{zhang2016efficient}, they do not consider load balancing, which limits the improvement of efficiency.

\vs
\noindent
\underline{\textbf{Contributions.}}
Motivated by the above facts, we present parallel DPC algorithms that run in $o(n^2)$ time under a reasonable assumption.
We summarize, in Table \ref{table_time-complexity}, the worst time complexities of existing and our algorithms, and our main contributions below\footnote{This is an error-corrected version of \cite{amagata2021fast}.
Note that the algorithms do not change, so their practical performances also keep efficient.}.
\begin{itemize}
	\setlength{\leftskip}{-4.0mm}
        \item	\underline{\textit{Ex-DPC} (\S \ref{section_exdpc})}.
        	We first propose Ex-DPC, an exact algorithm that exploits a $k$d-tree \cite{bentley1975multidimensional}.
			This algorithm computes the local densities and dependent distances of all points in $O(n(n^{1-1/d} + \rho_{avg}))$ time and $O(n^{2-1/d})$ time, respectively.
        \item	\underline{\textit{Approx-DPC} (\S \ref{section_approxdpc})}.
        	Second, we propose Approx-DPC, an approximation algorithm that relaxes the constraint of dependent points \textit{by approximating dependent points}.
                This algorithm improves the performance of local density computation and computes approximate dependent points for many points in $O(1)$ time.
                Although the worst time complexity of Approx-DPC is the same as that of Ex-DPC, its time is shown to be faster than $O(n(n^{1-1/d} + \rho_{avg}))$ under a practical assumption.
	\item	\underline{\textit{S-Approx-DPC} (\S \ref{section_sapproxdpc})}.
        	Our last algorithm, S-Approx-DPC, is also an approximation algorithm.
                S-Approx-DPC employs the concepts of \textit{grid sampling} and \textit{cell-based clustering} so that users can improve computation time through an approximation parameter.
                We theoretically discuss the efficiency of S-Approx-DPC.
	\item	\underline{\textit{Parallelization of our algorithms}}.
			We also present how to parallelize our algorithms in a multicore environment.
	\item	\underline{\textit{Empirical evaluation} (\S \ref{section_experiment})}.
        	We conduct experiments using synthetic and real datasets.
                Our experimental results demonstrate that Ex-DPC outperforms the state-of-the-art exact algorithm significantly.
                Also, our approximation algorithms beat the state-of-the-art approximation algorithms and are usually more than 10 times faster than them.
\end{itemize}
We formally define the problem in this paper and introduce related works in \S \ref{section_preliminary}, and this paper is concluded in \S \ref{section_conclusion}.

\section{Preliminary}	\label{section_preliminary}

\subsection{Problem Definition} \label{section_problem-definition}
Let $P$ be a set of $n$ points in a $d$-dimensional space $\mathbb{R}^d$.
Density-Peaks Clustering (DPC) aims at dividing $P$ into some subsets based on density-peaks.
To this end, DPC requires two important metrics, \textit{local density} and \textit{dependent distance}.
We formally define them.

\begin{defi}[\textsc{Local density}]	\label{definition_local-density}
Given a point $p_{i} \in P$ and a cutoff distance $d_{cut}$, its local density $\rho_{i}$ is:
\begin{equation}
	\rho_{i} = |\{p_{j} \,|\, dist(p_{i},p_{j}) < d_{cut}, p_{j} \in P\}|. \label{eq:rho}
\end{equation}
\end{defi}

\noindent
That is, the local density of $p_{i} \in P$ is the number of points $p_{j}$ such that $dist(p_{i},p_{j})$ is less than a user-specified threshold $d_{cut}$\footnote{The analyses in \cite{amagata2021fast} are essentially valid iff the local density is defined as an axis-parallel rectangle.
This paper adds some conditions to make the analyses still valid for the circular (or hyper-ball) case.}.

\begin{defi}[\textsc{Dependent point}]	\label{definition_dependent-point}
\textit{Given a point $p_{i} \in P$, its dependent point $q_{i}$ satisfies:}
\begin{equation*}
	q_{i} = \argmin_{\rho_{i} < \rho_{j}} dist(p_{i},p_{j}).
\end{equation*}
\end{defi}

\noindent
That is, the dependent point of $p_{i}$ is the nearest neighbor point to $p_{i}$ among a set of points with higher local densities than $\rho_{i}$ (ties are broken arbitrarily).

\begin{defi}[\textsc{Dependent distance}]	\label{definition_dependent-distance}
\textit{Given a point $p_{i}$, its dependent distance $\delta_{i}$ is $dist(p_{i},q_{i})$.}
\end{defi}

\noindent
The point with the highest local density in $P$ cannot have a dependent point, so its dependent distance is simply $\infty$.

Next, we define \textit{noise} and \textit{cluster center}, which are respectively based on local density and dependent distance.

\begin{defi}[\textsc{Noise}]
\textit{If a point $p_{i} \in P$ has $\rho_{i} < \rho_{min}$, $p_{i}$ is a noise.}
\end{defi}

\begin{defi}[\textsc{Cluster center}]
\textit{If a non-noise point $p_{i} \in P$ has $\delta_{i} \geq \delta_{min}$, where $\delta_{min} > d_{cut}$ is a user-specified threshold}, $p_{i}$ is a cluster center.
\end{defi}

\noindent
We can specify $\rho_{min}$ and $\delta_{min}$ at the same time when $d_{cut}$ is specified or after a decision graph is viewed.
Note that $\rho_{min}$ is specified to remove points with (very) small local densities (e.g., $\rho_{min} = 10$).
Besides, $\delta_{min}$ is specified so that points with much longer dependent distances than the other points (like ones in Figure \ref{fig_decision-graph}) are selected as cluster centers.

Once a cluster center, say $p_{i}$, is identified, we consider that the dependent point of $p_{i}$ is itself.
There are points $p_{j}$ whose dependent points are $p_{i}$.
Also, there are points $p_{k}$ whose dependent points are $p_{j}$.
From this observation, we say that $p_{k}$ (and also $p_{i}$ and $p_{j}$) is (are) \textit{reachable} from the cluster center $p_{i}$.
Based on this, we define:

\begin{defi}[\textsc{Cluster}]
\textit{Given $P$, a cluster $C$, whose cluster center is $p_{i}$, is a non-empty subset of $P$ such that non-noise points $p \in C$ are reachable from $p_{i}$.}
\end{defi}

\noindent
Since each point $p \in P$ has a single dependent point, $p$ belongs to a single cluster.
Therefore, DPC provides a unique set of clusters.

We assume that $P$ is memory-resident and in a low-dimensional Euclidean space \cite{gan2015dbscan, gan2017dynamic, gan2017hardness, gan2018fast, song2018rp, wang2020theoretically}.
For high-dimensional data, dimensionality reduction, e.g., \cite{mcinnes2018umap}, is commonly used to obtain meaningful clusters \cite{aggarwal2013data}.
Therefore, our assumption fits into practical usage.
In addition, we put a practical assumption that $d_{cut}$ is sufficiently small to satisfy $\rho_{avg} \ll n$ (or $\rho_{avg} < n$), where $\rho_{avg}$ is the average local density in $P$.
If $\rho_{avg} \approx n$, we cannot see density-peaks, because most points in $P$ have very large local densities.
This case hence renders a bad clustering result.
Last, this paper assumes a single machine with a multicore CPU (or with multicore CPUs).
The other parallel computation environments are not the scope of this paper.

\subsection{Straightforward algorithm}
One simple solution of DPC is as follows.
Given a set $P$ of points,
\begin{enumerate}
	\setlength{\leftskip}{-3.0mm}
	\item	For each point $p_{i} \in P$, we compute its local density $\rho_{i}$ by a linear scan of $P$.
    \item	We sort $P$ in descending order of local density.
    \item	Then, for each point $p_{i} \in P$, we compute its dependent point $q_{i}$ through a linear scan of $P$ with early termination (we can stop the scan when we access $p_{j}$ such that $\rho_{i} \geq \rho_{j}$).
        	To efficiently process the next operation, each point $p_{i}$ in $P$ maintains the identifiers of points whose dependent points are $p_{i}$ (e.g., if $p_{i}$ is $q_{j}$, $p_{i}$ maintains $j$).
    \item	Last, we determine noises and cluster centers and then propagate the corresponding cluster label for each non-noise point in the manner of depth-first-search from the cluster centers.
\end{enumerate}

The above operations incur $O(n^{2})$, $O(n\log n)$, $O(n^{2})$, and $O(n)$ time, respectively.
The above algorithm hence incurs $O(n^{2})$ time.
This is not tolerant of large datasets, thus we devise efficient solutions.
The above label propagation operation is already efficient and common to our algorithms.

\subsection{Related Work}	\label{section_related-work}
\noindent
\underline{\textbf{$k$-means clustering}} is to find a set of clusters that minimizes the sum of squared deviations of points in the same cluster.
This is well-known to be NP-hard, and many studies developed fast approximation algorithms with a small error \cite{arthur2007k, bahmani2012scalable}.
One limitation of $k$-means clustering is that it normally provides ball-shaped clusters and cannot deal with complex-shape clusters.
Another limitation is that $k$-means clustering is sensitive to noises (or outliers).
Some papers addressed the problem of $k$-means clustering in the presence of noises \cite{ceccarello2019solving, gupta2017local}.
However, this problem requires the number of noises in advance, which is not practical.

\vs
\noindent
\underline{\textbf{DBSCAN.}}
In this clustering, each point $p \in P$ is evaluated whether it is a \textit{core point} or not, based on two input parameters $\varepsilon$ and $minPts$ (if at least $minPts$ points exist within $\varepsilon$ from $p$, $p$ is a core point) \cite{ester1996density}.
DBSCAN assumes that, if the distance between two core points is within $\varepsilon$, there is a connection between them.
Informally, DBSCAN forms a cluster by connecting core points in the above way.

We do not say that DPC can replace DBSCAN, since an appropriate clustering algorithm for a given dataset is dependent on its data distribution.
However, DPC is more effective for datasets that have skewed density and points existing between close clusters.
This is because DBSCAN may consider multiple dense point groups as a single cluster if there are points existing in the border spaces between different groups, while DPC is robust to such a data distribution, as shown in Figure \ref{fig_s2-dbscan}.
This is the main difference between DPC and DBSCAN\footnote{A further discussion about their difference can be found in \cite{gong2017clustering}.
Besides, \cite{bai2017fast, gong2017clustering, ulanova2016clustering, zhang2016efficient} compared DPC and DBSCAN w.r.t. clustering qualities, and the results are similar to Figure \ref{figure_s2-cluster}, i.e., clustering results are (totally) different.
Interested readers may refer to them.}.
Variants of DBSCAN, such as OPTICS \cite{ankerst1999optics}, inherit the above problem \cite{bai2017fast} (see Example 2) and do not scale to large datasets, because they incur $O(n^2)$ time.

DBSCAN is costly, so efficient stand-alone algorithms \cite{chen2018fast, schubert2017dbscan} and parallel ones \cite{lulli2016ng, yang2019dbscan} have been proposed.
Due to the computational hardness of exact DBSCAN \cite{gan2015dbscan, gan2017hardness}, approximation algorithms for DBSCAN have been receiving attention \cite{gan2015dbscan, gan2017hardness, song2018rp, wang2019dbsvec}.

\vs
\noindent
\underline{\textbf{Density-peaks clustering.}}
Many applications have employed DPC, and some variants of DPC \cite{chen2020fast, mehmood2017clustering, yang2019streamline, wang2016automatic} and streaming DPC \cite{gong2017clustering, ulanova2016clustering} have also been proposed.
This paper follows the original definition \cite{rodriguez2014clustering} and considers a static $P$.

One state-of-the-art parallel and approximation DPC algorithm is LSH-DDP \cite{zhang2016efficient}.
Although LSH-DDP is originally designed for distributed computing environments (MapReduce), it can work in multicore environments.
LSH-DDP employs locality-sensitive hashing (LSH) \cite{datar2004locality} to partition $P$ into some buckets (i.e., disjoint subsets of $P$) so that points in the same bucket are similar to each other.
For each $p \in P$, LSH-DDP computes an approximate local density of $p$ by scanning the bucket that includes $p$.
Then, an approximate dependent point of $p$ is retrieved from the bucket that includes $p$.
If the distance between $p$ and its approximate dependent point does not seem accurate, LSH-DDP computes its dependent point by scanning $P$.
LSH-DDP can reduce computation time by approximating local densities and dependent distances, but it does not consider load balancing for parallel processing.

One state-of-the-art exact algorithm for DPC is CFSFDP-A \cite{bai2017fast}.
This algorithm selects \textit{pivot points} and utilizes triangle inequality to avoid unnecessary distance computation.
In the pre-processing phase, CFSFDP-A employs $k$-means clustering to select pivot points, that is, $k$ points are selected as pivot and they are the centroids of $k$ clusters.
For each point $p_{i} \in P$, CFSFDP-A computes a candidate set of points $p_{j}$ such that $dist(p_{i},p_{j})$ may be less than $d_{cut}$ by utilizing pivots and triangle inequality.
Its dependent point computation is done in a similar manner.
Recall that $k$-means clustering is sensitive to noises, so its pivot selection does not provide good filtering power, meaning that the candidate size is still large.
An approximation algorithm was also proposed in \cite{bai2017fast}, but it does not consider density-peaks and provides clustering results with low accuracy.

\section{Ex-DPC}	\label{section_exdpc}
Ex-DPC, our exact DPC algorithm, assumes that a point set $P$ is indexed by an in-memory $k$d-tree $\mathcal{K}$, which provides efficient tree update and similarity search in practice while retaining theoretical performance guarantees\footnote{Ex-DPC (or our algorithms) can in fact employ a similar data structure, e.g., R-tree \cite{qi2018theoretically}.
However, we do not consider the R-tree for our solution in this paper because it does not provide any theoretical performance guarantee.}.
First, we present Ex-DPC with a single thread.
We then explain how to parallelize Ex-DPC.

\vs
\noindent
\underline{\textbf{Local density computation.}}
Definition \ref{definition_local-density} suggests that computing the local density of a point $p_{i} \in P$ corresponds to doing a range search whose query point and radius are $p_{i}$ and $d_{cut}$, respectively.
Because the $k$d-tree supports efficient range search, we simply utilize it.
That is, for each $p_{i} \in P$, Ex-DPC runs a range search on $\mathcal{K}$ to obtain $\rho_{i}$.
This approach is simple but has good theoretical performance.

Before we analyze the performance of Ex-DPC, we put the following assumption.

\begin{assumption}  \label{assumption:out}
For each point $p \in P$, let OUT be the number of points in the bounding rectangle of the circle defined in Equation (\ref{eq:rho}).
We assume that OUT $\approx \rho$.
\end{assumption}

\noindent
This is reasonable and practical for small $d_{cut}$.
Hereinafter, our theoretical analyses assume that Assumption \ref{assumption:out} holds, i.e., (some of) the subsequent lemmas and theorems hold under Assumption \ref{assumption:out}.

\begin{lemm}	\label{lemma_exdpc_local-density}
The time complexity of the local density computation in Ex-DPC is $O(n(n^{1-1/d} + \rho_{avg}))$ under Assumption \ref{assumption:out}.
\end{lemm}

\noindent
\textsc{Proof.}
The time complexity of a range search with query point $p_{i}$ on a $k$d-tree is $O(n^{1-1/d} + \rho_{i})$ \cite{toth2017handbook}.
Therefore, the time complexity of the local density computation of Ex-DPC is $O(\sum_{P}(n^{1-1/d} + \rho_{i})) = O(n(n^{1-1/d} + \rho_{avg}))$.	\wsq

\vs
\noindent
\underline{\textbf{Dependent point computation.}}
Recall the constraint of dependent points: the dependent point of $p_{i}$ has to be retrieved from a set of points with higher local densities than $\rho_{i}$.
Since the local density depends on $d_{cut}$, it is hard to build a data structure for obtaining dependent points efficiently in a pre-processing phase.
Although the $k$d-tree supports efficient nearest neighbor search, it is not guaranteed that the nearest neighbor point of $p_{i}$ has higher local density than $\rho_{i}$.
Hence it is challenging to compute dependent points efficiently.
We overcome this challenge with an idea that \textit{an optimal $k$d-tree for computing $q_{i}$ can be built incrementally}.

Ex-DPC computes the dependent point of each point in $P$ in the following way:
\begin{enumerate}
    \setlength{\leftskip}{-3.0mm}
    \item	Destroy $\mathcal{K}$ (i.e., $\mathcal{K}$ becomes an empty set).
    \item	Sort $P$ in descending order of local density.
    \item	Pick the front point of $P$, say $p$, set $\infty$ as its dependent distance, insert $p$ into $\mathcal{K}$, and pop $p$.
    \item	Pick the front point of $P$, say $p'$, conduct a nearest neighbor search with query point $p'$ on $\mathcal{K}$, set the result as its dependent point, insert $p'$ into $\mathcal{K}$, and pop $p'$.
    \item	Repeat the above operation until $P$ has no points.
\end{enumerate}

It is important to note that, for the front point $p_{i}$ of $P$, the $k$d-tree contains only points whose local densities are higher than $\rho_{i}$.
(We assume that all points have different local densities, which is practically possible by adding a random value $\in (0,1)$ to $\rho_{i}$.)
Therefore, for $p_{i}$, its nearest neighbor search retrieves the correct dependent point.
We now prove that our approach is efficient.

\begin{lemm}	\label{lemma_exdpc_dependent-point}
Ex-DPC needs $O(n(n^{1-1/d}+\rho_{avg}))$ time to compute the dependent points of all points in $P$, when most points in $P$ have $\delta \leq d_{cut}$ (i.e., when we have $\alpha(n-1)(n^{1-1/d} + \rho_{avg}) > (1-\alpha)(n-1)n$ for $\alpha \in [0,1]$).
\end{lemm}

\noindent
\textsc{Proof.}
Destroying $\mathcal{K}$ and sorting $P$ incur $O(n\log n)$ time, and inserting a point into an empty $\mathcal{K}$ incurs $O(1)$ time.
Next, for a point $p \in P$, computing its dependent point incurs $O(n^{1-1/d} + \rho_{i})$ time in the worst case if $\delta_{i} \leq d_{cut}$, which is seen from the range search result.
Inserting $p_{i}$ into $\mathcal{K}$ incurs $O(\log n)$ time.
Therefore, the fourth operation needs $O(n^{1-1/d}+\rho_{i})$ time if $\delta_{i} \leq d_{cut}$ (otherwise, it needs $O(n)$ time).
Let $\alpha$ be the ratio that a given $p \in P$ has $\delta \leq d_{cut}$, i.e., $\alpha n$ points have this case.
Since we repeat the above operation $(n - 1)$ times, our approach incurs $O(\alpha(n-1)(n^{1-1/d} + \rho_{avg}) + (1-\alpha)(n-1)n)$ time.
When most points in $P$ have $\delta \leq d_{cut}$, i.e., $\alpha$ is sufficiently large, $\max\{\alpha(n-1)(n^{1-1/d} + \rho_{avg}), (1-\alpha)(n-1)n\} = \alpha(n-1)(n^{1-1/d} + \rho_{avg})$.
Therefore, this lemma holds.  \wsq

\vs
\noindent
In practice, most points $\in P$ have $\delta \leq d_{cut}$, thus this assumption is not strict.

\vs
\noindent
\underline{\textbf{Analysis.}}
From Lemmas \ref{lemma_exdpc_local-density} and \ref{lemma_exdpc_dependent-point} and the time complexity of the label propagation, we have:

\begin{theo}[\textsc{Time complexity of Ex-DPC}]
The time complexity of Ex-DPC is $O(n(n^{1-1/d} + \rho_{avg}))$, when most points in $P$ have $\delta \leq d_{cut}$.
\end{theo}

\noindent
Now we see that, for $d_{cut}$ yielding $n \cdot \rho_{avg} = o(n^2)$, Ex-DPC always needs time less than $O(n^2)$, and an arbitrary sufficiently small $d_{cut}$ satisfies it\footnote{Actually, as long as $\rho_{avg} \leq n^{1-1/d}$, Ex-DPC is clearly sub-quadratic to $n$.
As we assume small $d_{cut}$, this holds (for Ex-DPC, Approx-DPC, and S-Approx-DPC) in practice.}.
The space complexity of Ex-DPC is $O(n)$, as Ex-DPC employs a single $k$d-tree, whose space complexity is $O(n)$.

\vs
\noindent
\underline{\textbf{Implementation for parallel processing.}}
We can parallelize the local density computation in Ex-DPC, but, unfortunately, its dependent point computation cannot be parallelized.
This is derived from the fact that Ex-DPC needs to compute the dependent point of each point \textit{one by one}, since the $k$d-tree is incrementally updated.
Hence we focus on how to parallelize its local density computation.

Given $P$ and multiple threads (or CPU cores), we parallelize local density computation by assigning each point in $P$ to one of the threads.
Then, each thread independently conducts a range search for each assigned point.
To exploit the parallel processing environment (i.e., to hold load balancing), each thread should have (almost) the same processing cost.
Recall that the range search cost of $p_{i}$ is $O(n^{1-1/d} + \rho_{i})$, indicating that the cost depends on its local density, which cannot be pre-known and is different between points.
This means that a simple hash-partitioning of $P$ may not hold load balancing.
We therefore employ a heuristic that assigns a point to a thread \textit{dynamically}.
Specifically, for each thread, Ex-DPC assigns a point, and when a thread has finished its range search, Ex-DPC assigns another point to the thread.
We use OpenMP for multi-threading, and to implement the above approach, ``\textsf{\#pragma omp parallel for schedule (dynamic)}'' is used.

\section{Approx-DPC}	\label{section_approxdpc}
Ex-DPC still has some weaknesses.
(i) Ex-DPC incurs unnecessary $k$d-tree traversal, because points with small distances traverse almost the same nodes of $\mathcal{K}$.
(ii) Ex-DPC has the hardness of parallelizing its dependent point computation.
Our first approximation DPC algorithm, Approx-DPC, removes the above limitations to accelerate processing efficiency by joint range search and approximating dependent points.
We prove that Approx-DPC provides the same cluster centers, given the same $\rho_{min}$ and $\delta_{min}$ for Ex-DPC, rendering an accurate clustering result.

\subsection{Data Structure}	\label{section_approxdpc-data-structure}
Approx-DPC also uses a $k$d-tree $\mathcal{K}$ to index $P$.
In addition, Approx-DPC leverages another data structure, which is a uniform grid $G$, a set of non-empty cells, where each cell is a $d$-dimensional square with side length $\frac{d_{cut}}{\sqrt{d}}$ for all dimensions.
Each cell $c$ of the grid $G$ maintains:
\begin{itemize}
    \setlength{\leftskip}{-4.0mm}
    \item	$P(c)$: a set of points covered by $c$,
    \item	$p^{*}(c)$: the point with the maximum local density among $P(c)$,
    \item	$\min_{P(c)}\rho$: the minimum local density in $P(c)$, and
    \item	$N(c)$: an identifier set of cells to which points $p \notin P(c)$ satisfying $dist(p^{*}(c),p) < d_{cut}$ belong.
\end{itemize}
An example of the above grid structure is illustrated in Figure \ref{figure_grid}.

Approx-DPC builds the grid $G$ online, as it depends on $d_{cut}$.
Given $P$, we sequentially access each point $p \in P$ and map it to its corresponding cell.
If the cell has not been created, we create it before $p$ is mapped, thereby no empty-cell is created.
The information maintained in each cell is obtained during the local density computation.

Some existing clustering studies, e.g., \cite{gan2015dbscan, song2018rp}, also employ grid-based data structures, but the details are totally different.
It is important to note that our main contributions in \S \ref{section_approxdpc} are new ideas (joint range search and cell-based dependent point approximation) which respectively derive efficient local density and dependent point computation, through our grid $G$.

\subsection{Local Density Computation}	\label{section_approxdpc-local-density}
\begin{figure}[!t]
    \centering
    \includegraphics[width=0.99\linewidth]{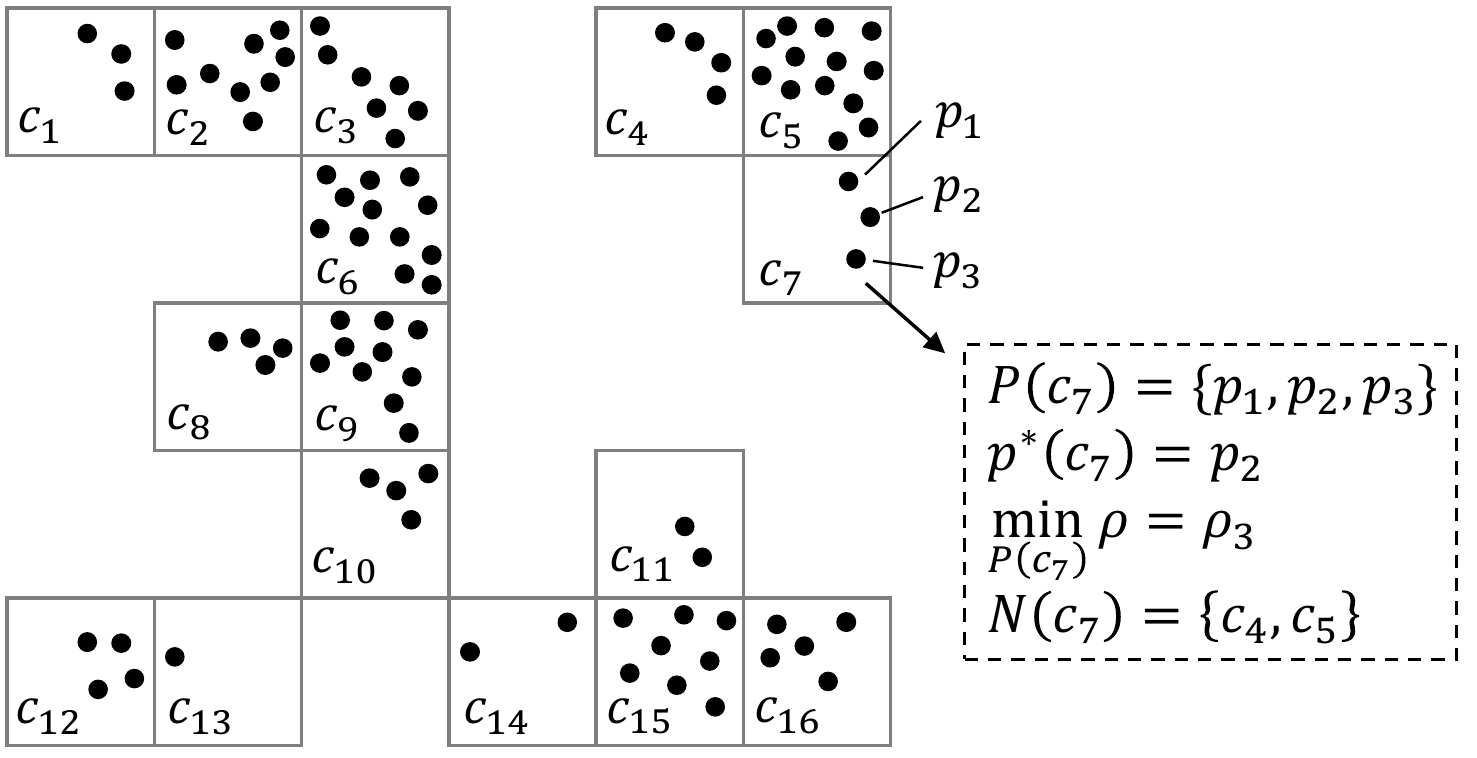}
    \caption{The grid structure in Approx-DPC.
    The black points represent points in $P$, and there are 16 cells in $G$ for a given $d_{cut}$.
    The dashed box shows each element maintained in cell $c_{7}$ where $\rho_{2}$ is the largest among $P(c_{7}) = \{p_{1},p_{2},p_{3}\}$.}
    \label{figure_grid}
\end{figure}

Let $B(p_{i},d_{cut})$ be the (open) ball centered at $p_{i}$ with radius $d_{cut}$.
If $dist(p_{i},p_{j})$ is small, $B(p_{i},d_{cut})$ and $B(p_{j},d_{cut})$ have a significant overlap.
For instance, as can be seen in Figure \ref{figure_grid}, points in the same cell have this observation.
This suggests that Ex-DPC incurs unnecessary $k$d-tree traversal (recall that Ex-DPC iteratively conducts a range search).
Approx-DPC improves the performance of local density computation by using an idea: \textit{if we can jointly search the points covered by each ball $B(p,d_{cut})$, where $p \in P(c)$, in a cell $c$, we can reduce unnecessary tree traversal}.
For $p \in P$, Approx-DPC computes its exact local density, to guarantee that the same cluster centers are selected as those in Ex-DPC.

\begin{figure}[!t]
    \centering
    \includegraphics[width=0.99\linewidth]{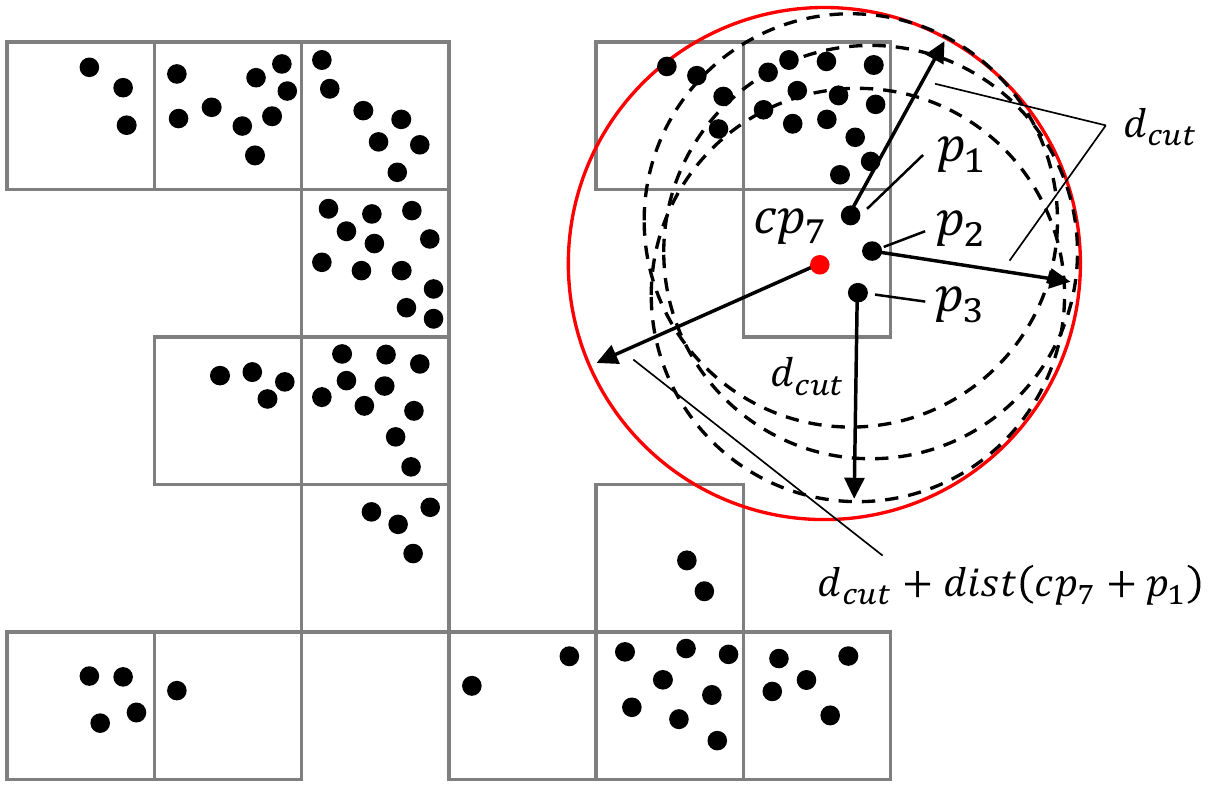}
    \caption{An example of joint range search}
    \label{figure_joint-search}
\end{figure}

\vs
\noindent
\underline{\textbf{Joint range search.}}
Given a cell $c_{i}$ of $G$, we obtain points covered by each ball of a point $p \in P(c_{i})$, through a single range search on $\mathcal{K}$.
Let $cp_{i}$ be the center point of $c_{i}$ (i.e., the coordinates of $cp_{i}$ are the center of $c_{i}$).
Furthermore, define $p' = \argmax_{p \in P(c_{i})}dist(cp_{i},p)$.
It is important to observe that, for each $p \in P({c_{i}})$, we have $B(p,d_{cut}) \subseteq B(cp_{i},d_{cut} + dist(cp_{i},p'))$.
This corresponds to the following: let $R(p,d_{cut})$ be the result of the range search with query point $p$ and radius $d_{cut}$.
Then $R(cp_{i},d_{cut} + dist(cp_{i},p'))$ is a superset of $R(p,d_{cut})$ for each $p \in P(c_{i})$.
Figure \ref{figure_joint-search} depicts this observation, where the red point is the center point of cell $c_{7}$.

We proceed to present the procedure of the joint range search in a cell.
Given a cell $c_{i}$ of $G$, Approx-DPC
\begin{enumerate}
	\setlength{\leftskip}{-3.0mm}
    \setlength{\itemsep}{0.0mm}
    \item	makes $cp_{i}$ (the center point of $c_{i}$),
    \item	computes $\max_{p \in P(c_{i})} dist(cp_{i},p)$ (assume that $p' \in P(c_{i})$ holds this),
    \item	obtains $R(cp_{i},d_{cut} + dist(cp_{i},p'))$ through a range search with query point $cp_{i}$ and radius $d_{cut} + dist(cp_{i},p')$ on $\mathcal{K}$, and
    \item	for each $p_{j} \in P(c_{i})$, scans $R(cp_{i},d_{cut} + dist(cp_{i},p'))$ to compute its exact local density $\rho_{j}$.
\end{enumerate}

\noindent
\underline{\textbf{Algorithm.}}
Given a cell $c_{i}$ of $G$, Approx-DPC conducts a joint range search.
During this, Approx-DPC computes the point with the maximum local density in $c_{i}$, $p^{*}(c_{i})$, and the minimum local density in $c_{i}$, $\min_{P(c_{i})}\rho$.
When $p^{*}(c_{i})$ is identified, $N(c_{i})$, the identifier set of cells $c_{j}$ that cover $p \notin P(c_{i})$ such that $dist(p^{*}(c_{i}),p) < d_{cut}$, is obtained.
Approx-DPC repeats this operation for every cell of $G$.

\vs
\noindent
\underline{\textbf{Time complexity.}}
Let $\rho(cp_{i}) = |R(cp_{i},d_{cut} + dist(cp_{i},p'))|$.

\begin{lemm}	\label{lemma_approxdpc_local-density}
Approx-DPC needs $O(\sum_{G}(n^{1-1/d} + \rho(cp_{i}) \cdot |P(c_{i})|))$ time to compute the local densities of all points.
\end{lemm}

\noindent
\textsc{Proof.}
Obviously, the main cost of the local density computation in Approx-DPC is derived from joint range searches, because the cell update (i.e., computing $p^{*}(c_{i})$, $\min_{P(c_{i})}\rho$, and $N(c_{i})$) is done during the joint range search.
The joint range search in a cell $c_{i}$ consists of (i) the range search with query point $cp_{i}$ and radius $d_{cut} + dist(cp_{i},p')$ on the $k$d-tree and (ii) exact local density computation for each $p \in P(c_{i})$ by scanning the range search result $R(cp_{i},d_{cut} + dist(cp_{i},p'))$.
The cost of (i) is $O(n^{1-1/d} + \rho(cp_{i}))$, and the cost of (ii) is $\Theta(\rho(cp_{i}) \cdot |P(c_{i})|)$.
The joint range search time in $c_{i}$ is then $O(n^{1-1/d} + \rho(cp_{i}) \cdot |P(c_{i})|)$.
Consequently, we have the lemma.	\wsq

\begin{remark}  \label{remark_approx-dpc}
The local density computation in Approx-DPC beats the one in Ex-DPC in practice.
For small $d_{cut}$, we have $\rho(cp_{i}) = O(\rho_{j})$ in practice, where $p_{j} \in P(c_{i})$.
Let $|P(c_{i})| = n_{i}$.
We then have
\begin{align*}
	O(\sum_{G}(n^{1-1/d} + \rho(cp_{i}) \cdot n_{i})) &	= O(\sum_{G}n^{1-1/d} + \sum_{G}\rho_{j} \cdot n_{i})	\\
    &	= O(\sum_{G}n^{1-1/d} + \sum_{P}\rho_{j}).
\end{align*}
Because $|G| \leq n$, $\sum_{G}n^{1-1/d} \leq n^{2-1/d}$.
Moreover, $\sum_{P}\rho_{j} = n \cdot \rho_{avg}$.
We now see the improvement of local density computation against the one in Ex-DPC.
\end{remark}

\subsection{Dependent Point Computation}	\label{section_approxdpc-dependent-point}
Ex-DPC demonstrates that, for each point in $P$, its dependent point is obtained efficiently, but parallelizing it is hard.
A challenge here is how to compute dependent points while holding efficiency and parallelizability.
We solve this challenge by allowing \textit{approximate dependent points}, and design a fast algorithm based on the following key ideas:
(i) Clustering accuracy would not degrade as long as
\begin{itemize}
    \setlength{\leftskip}{-5.0mm}
    \item	an approximate dependent point of each point $p_{i}$ is close to $p_{i}$, and
    \item	we compute the exact dependent point for $p_{i}$ if there are no close points with higher local density than $\rho_{i}$.
\end{itemize}
(ii) It is easy to find an approximate dependent point of $p_{i}$ if we use the information maintained in each cell of $G$.

\vs
\noindent
\underline{\textbf{Approximate computation.}}
To implement the above key ideas, Approx-DPC allows the following: \textit{for a point $p_{i} \in P$, if there exists a point $p_{j}$ such that $\rho_{i} < \rho_{j}$ and $dist(p_{i},p_{j}) \leq d_{cut}$, $p_{j}$ can be an approximate dependent point of $p_{i}$.}
Due to this, many points in $P$ can have their approximate dependent points in a constant time.
Specifically, Approx-DPC computes an approximate dependent point of $p_{i}$, which belongs to a cell $c$, based on the following rules:
\begin{itemize}
    \setlength{\leftskip}{-4.0mm}
    \item	If $p_{i} \neq p^{*}(c)$, its approximate dependent point is $p^{*}(c)$,
        	(The distance to each point in the same cell is at most $d_{cut}$.)
        	The dependent distance of $p_{i}$ is set as $d_{cut}$, instead of computing the distance between $p_{i}$ and its approximate dependent point.
            (We later show that it does not matter.)
    \item	If $p_{i} = p^{*}(c)$, Approx-DPC retrieves a cell $c'$ from $N(c)$ such that $\min_{P(c')}\rho > \rho_{i}$.
        	If there exists such a cell $c'$, the approximate dependent point and dependent distance of $p_{i}$ are $p^{*}(c')$ and $d_{cut}$, respectively.
                Otherwise, we do not decide its approximate dependent point here.
\end{itemize}
It is important to note that $|N(c)| = O(1)$ for an arbitrary fixed $d$, meaning that the above approach takes only $O(1)$ time for each point $p \in P$.

\vs
\noindent
\underline{\textbf{Exact computation.}}
For a point $p_{i}$ whose dependent point has not been decided in the above computation, Approx-DPC computes its exact dependent point.
We take a different approach from Ex-DPC, and our approach in Approx-DPC still runs in $o(n^2)$ time and holds parallelizability.
Our idea here is as follows:
For $p_{i}$, we avoid accessing points with less local density than $\rho$ by dividing $P$ into some subsets and retrieve $q_{i}$ by pruning unnecessary subsets.

\begin{figure}[!t]
    \centering
    \includegraphics[width=0.99\linewidth]{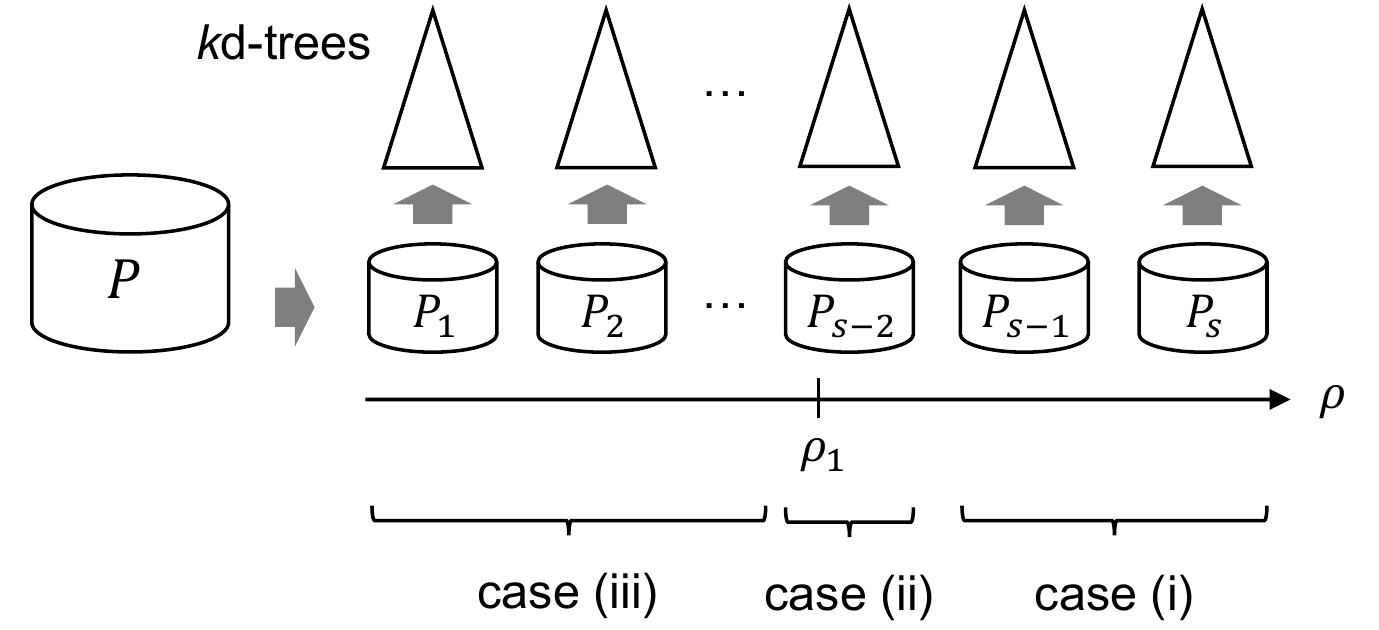}
    \caption{Partitioning $P$ into $s$ disjoint subsets, and the relationship between a point $p_{1}$ and the subsets}
    \label{figure_partition}
\end{figure}

First, we sort $P$ in ascending order of local density.
Second, we equally divide $P$ into $s$ disjoint subsets, $P_{1}$, ..., $P_{s}$ (i.e., $P_{i} \cap P_{j} = \varnothing$ for $i \neq j$, and $\bigcup_{s}P_{i} = P$), and build a $k$d-tree $\mathcal{K}_{i}$ for each subset $P_{i}$.
Note that, for $p \in P_{i}$ and $p' \in P_{j}$ ($i < j$), we have $\rho < \rho'$.
We set $s$ so that
\begin{equation}
	\frac{n}{s} = O((s - 1)(\frac{n}{s})^{1-1/d})	\label{equation_s},
\end{equation}
to provide a theoretical performance guarantee (see Lemma \ref{lemma_approxdpc_dependent-point}).
Let $P'$ be the set of points whose dependent points have not been decided.
Given a point $p_{i} \in P'$ and a subset $P_{j}$, as can be seen from Figure \ref{figure_partition}, we have three cases:
\begin{enumerate}
    \renewcommand{\labelenumi}{(\roman{enumi})}
    \setlength{\leftskip}{-2.0mm}
    \item	All points in $P_{j}$ have higher local density than $\rho_{i}$.
		In this case, we conduct a nearest neighbor search on $\mathcal{K}_{j}$.
    \item	$P_{j}$ has not only points with higher local density than $\rho_{i}$ but also points $p_{k}$ that have $\rho_{i} \geq \rho_{k}$.
		In this case, we scan the whole $P_{j}$ and obtain the nearest neighbor point with higher local density than $\rho_{i}$ in $P_{j}$.
		Note that there is at most one subset which has this case for $p_{i}$.
    \item	The local densities of all points in $P_{j}$ are less than (or equal to) $\rho_{i}$.
		In this case, $p_{i}$ ignores $P_{j}$.
\end{enumerate}
The exact dependent point of $p_{i} \in P'$ is obtained by evaluating each subset $P_{j}$ based on the above approach.
The dependent distance of $p_{i}$, $\delta_{i}$, follows Definition \ref{definition_dependent-distance}.

\vs
\noindent
\underline{\textbf{Time complexity.}}
We analyze the efficiency of the dependent point computation in Approx-DPC.
Clearly, its main cost is derived from the exact dependent point computation, since the approximate computation requires only $O(1)$ time for each $p \in P$.
We thus consider the time complexity of the exact computation.

\begin{lemm}	\label{lemma_approxdpc_dependent-point}
The time complexity of the dependent point computation in Approx-DPC is $O(ns(\frac{n}{s})^{1-1/d})$.
\end{lemm}

\noindent
\textsc{Proof.}
Sorting $P$ needs $O(n\log n)$ time, and dividing $P$ into $s$ subsets needs $O(n)$ time.
Building a $k$d-tree for each subset requires $O(s \cdot \frac{n}{s}\log\frac{n}{s}) = O(n\log\frac{n}{s})$.
For a point $p_{i} \in P'$, its worst case is to conduct a nearest neighbor search on $\mathcal{K}_{2}$, ..., $\mathcal{K}_{s}$ and scan the whole $P_{1}$.
This case incurs $O((s-1)(\frac{n}{s})^{1-1/d})$ from Equation (\ref{equation_s}).
Since $|P'| = O(n)$, we have the lemma.	\wsq

\vs
\noindent
Equation (\ref{equation_s}) implies that $s$ never reaches $n$ (and $s$ is small in practice), so this lemma theoretically demonstrates the efficiency of the dependent point computation in Approx-DPC.
It is important to note that, since many points obtain their approximate dependent points in $O(1)$ time, meaning $|P'| \ll n$, its practical cost is much less than the one in Lemma \ref{lemma_approxdpc_dependent-point}.
In addition, we will clarify that this approach is parallel-friendly in \S \ref{section_approxdpc-parallel}.

\subsection{Analysis}	\label{section_approxdpc_analysis}
Next, we consider the overall performance of Approx-DPC.
Lemmas \ref{lemma_approxdpc_local-density} and \ref{lemma_approxdpc_dependent-point} and the fact that $|P'| \ll n$ prove the following theorem.

\begin{theo}[\textsc{Time complexity of Approx-DPC}]
Approx-DPC requires $O(\sum_{G}(n^{1-1/d} + \rho(cp_{i}) \cdot |P(c_{i})|))$ for an arbitrary fixed $d$.
\end{theo}

\noindent
Remark \ref{remark_approx-dpc} has already claimed that this time complexity is better than that of Ex-DPC in practice.
Besides, Approx-DPC has reasonable space complexity.

\begin{theo}[\textsc{Space complexity of Approx-DPC}]	\label{theorem_approxdpc_space}
\textit{The space complexity of Approx-DPC is $O(n)$ for an arbitrary fixed $d$.}
\end{theo}

\noindent
\textsc{Proof.}
Approx-DPC employs a $k$d-tree and a grid.
The $k$d-tree $\mathcal{K}$ requires $O(n)$ space, and $\bigcup P(c) = P$.
In addition, $|N(c)| = O(1)$ for an arbitrary fixed $d$.
The exact dependent point computation in Approx-DPC builds $s$ $k$d-trees, each of which has $O(\frac{n}{s})$ space, i.e., $O(n)$ in total.
This concludes that the theorem is true.	\wsq

\vs
\noindent
\underline{\textbf{Why accurate?}}
Approx-DPC provides a highly accurate clustering result, which is empirically demonstrated in \S \ref{section_experiment}.
We discuss why we have this.
To start with, it is important to see:

\begin{theo}[\textsc{Cluster center guarantee}]
\textit{Approx-DPC provides the same cluster centers as Ex-DPC, given the same $\delta_{min}$ and $\rho_{min}$.}
\end{theo}

\noindent
\textsc{Proof.}
For a point $p_{i} \in P$, Approx-DPC approximates its dependent distance, which is $d_{cut}$, iff there exists a point $p_{j}$ such that $\rho_{i} < \rho_{j}$ and $dist(p_{i},p_{j}) \leq d_{cut}$.
Recall that Approx-DPC computes the exact local density for all points in $P$ and finds the nearest neighbor point with higher local density for points $p_{i} \in P$ that do not have points $p_{j}$ satisfying $\rho_{i} < \rho_{j}$ and $dist(p_{i},p_{j}) \leq d_{cut}$.
This leads to that (1) noises are selected correctly and (2) Approx-DPC computes the \textit{exact dependent distance} for points $p_{i}$ with $\delta_{i} > d_{cut}$.
Since $\delta_{min} > d_{cut}$, this theorem holds.	\wsq

\vs
Assuming that there is a link (or an edge) between $p_{i}$ and $p_{j}$, where $p_{j} $ is the dependent point of $p_{i}$, a cluster in DPC is considered to be a tree rooted at a cluster center \cite{gong2017clustering}.
In this sense, a cell $c_{i}$ is a sub-tree rooted at $p^{*}(c_{i})$.
Also, points whose dependent distances are exactly computed (i.e., the roots of some sub-trees) are the stem of the tree (i.e., cluster).
The other sub-trees, which have approximate dependent points, are then considered to be branches connecting to (some parts of) the stem.
It is intuitively seen that, as long as the stem is exact and sub-trees are connected to it, the cluster is accurate.
Approx-DPC yields the exact stem and makes the other sub-trees connect to some part of the stem, thereby it forms an accurate cluster set.
One exception is border points that exist around the border between different clusters.
We study their influence in \S \ref{section_experiment}.

\subsection{Parallelization}	\label{section_approxdpc-parallel}
We parallelize Approx-DPC through a cost-based partitioning approach.
In a nutshell, given multiple threads, Approx-DPC (i) estimates the cost of each task in local density computation and dependent point computation and (ii) assigns the task to a thread, so that each thread has almost the same sum cost, for load balancing.
Although minimizing the sum cost difference between threads is NP-complete \cite{amagata2019identifying}, we can have a good partitioning result in practice through a 3/2-approximation greedy algorithm \cite{graham1969bounds}.
This greedy algorithm takes $O(n't)$ time, where $n'$ is the number of instances (cells or points) that are distributed to threads and $t$ is the number of threads.
This cost is trivial compared with those of the main operations, so we focus on how to estimate the cost of each task and how to parallelize Approx-DPC.

\vs
\noindent
\underline{\textbf{Parallel local density computation.}}
The joint range search of a cell $c_{i}$ consists of two main operations: obtaining $R(cp_{i},d_{cut} + dist(cp_{i},p'))$ and scanning it for each point in the cell.
The real cost of the first operation is hard to know in advance, since it depends on $d_{cut}$ and the density around $cp_{i}$.
However, we can estimate the density around $cp_{i}$ with a trivial cost.
Let $cost_{range}(c_{i})$ be the estimated cost of the range search with query point $cp_{i}$ and radius $d_{cut} + dist(cp_{i},p')$, and
\begin{equation*}
	cost_{range}(c_{i}) = |P(c_{i})|.
\end{equation*}
This is reasonable, because it clearly represents the density of the cell (if $|P(c_{i})|$ is large/small, $|R(cp_{i},d_{cut} + dist(cp_{i},p'))|$ should be large/small).
In addition, this estimation incurs a trivial computation cost, i.e., $O(1)$ time.
Next, let $cost_{scan}(c_{i})$ be the estimated cost of scanning $R(cp_{i},d_{cut} + dist(cp_{i},p'))$ for each point in $c_{i}$.
We have
\begin{equation*}
	cost_{scan}(c_{i}) = |P(c_{i})| \cdot |R(cp_{i},d_{cut} + dist(cp_{i},p'))|.
\end{equation*}
This estimation is possible in $O(1)$ time, after we obtain $R(cp_{i},d_{cut} + dist(cp_{i},p'))$.
We therefore employ a 2-phase approach.

We describe how to parallelize the local density computation in Approx-DPC.
Approx-DPC computes $cost_{range}(c_{i})$ for each cell $c_{i}$ in the grid.
Given multiple threads, Approx-DPC assigns $c_{i}$ to a thread, based on $cost_{range}(c_{i})$ and the greedy algorithm.
Each thread then independently conducts the range search and obtains $R(cp_{i},d_{cut} + dist(cp_{i},p'))$ for each assigned cell $c_{i}$.
Approx-DPC next estimates $cost_{scan}(c_{i})$ for each cell $c_{i}$, and again assigns $c_{i}$ to a thread, based on the same approach as the previous assignment.
Then, each thread independently computes the exact local densities of points in the assigned cells.

\vs
\noindent
\underline{\textbf{Parallel dependent point computation.}}
As for a point $p \in P$, Approx-DPC computes an approximate dependent point iff there exists a point that satisfies the approximation rules.
This is independent of computing approximate dependent points of the other points, so an approximate dependent point of each point can be computed in parallel.
Similarly, sorting $P$, dividing $P$ into $s$ subsets, and building a $k$d-tree for each subset can be also simply parallelized by assigning each point in $P$ or each subset $P_{i}$ into a given thread.
We hence consider the cost of computing the exact dependent point of $p \in P'$ ($P'$ is the set of points whose dependent points are exactly computed).

Now recall that its cost is dependent on the number of subsets $P_{i}$ which may have the dependent point of $p$.
Let $m$ be the number, and we estimate the cost as follows:
\begin{equation*}
	cost_{dep}(p) =
	\begin{cases}
        	\frac{n}{s} + (m - 1)(\frac{n}{s})^{1-1/d}	& (\text{if $p$ has case (ii)})	\\
                m(\frac{n}{s})^{1-1/d}					& (\text{otherwise}),
    \end{cases}
\end{equation*}
where $cost_{dep}(p)$ is an estimated cost.
We parallelize the exact dependent point computation by assigning $p \in P'$ to a thread based on $cost_{dep}(p)$ and the greedy algorithm.

\section{S-Approx-DPC}	\label{section_sapproxdpc}
Approx-DPC improves the performance of DPC against Ex-DPC without any additional parameters.
However, if applications allow a rough clustering result but require to obtain it more quickly, another solution is needed.
To this end, we propose S-Approx-DPC, which incorporates the following additional idea.

\vs
\noindent
\underline{\textbf{Main idea.}}
Given $P' = \{p_{i}, ...\} \subset P$, where points in $P'$ are close to each other, it is important to observe that points in $P'$ have almost the same local density.
Therefore their dependent points are the same or exist in a close area.
From this observation, we can consider that the clustering result does not change much, even if we pick only one point, say $p_{i}$, from $P'$, set $p_{i}$ as the (approximate) dependent point of the other points $p_{j}$ in $P'$, and do nothing for $p_{j}$.
(For S-Approx-DPC, $\rho_{min}$ is not applicable to $p_{j}$.)
Conceptually, S-Approx-DPC reduces the processing time by converting point clustering into cell clustering.
This reduces the number of range searches and the time to retrieve dependent points.

\vs
\noindent
\underline{\textbf{Data structure.}}
S-Approx-DPC also assumes that $P$ is indexed by a $k$d-tree, and builds a grid $G'$ online.
Each cell of $G'$ is a $d$-dimensional square with side length $\frac{\epsilon \cdot d_{cut}}{\sqrt{d}}$, where $\epsilon$ ($> 0$) is a user-specified approximation parameter.
Hereafter, we use $c$ to denote a cell of $G'$.
Different from the cells in Approx-DPC, each cell $c$ of $G'$ does not maintain $p^{*}(c)$ and $\min_{P(c)}\rho$, and $N(c)$ is defined as an identifier set of cells to which points $p' \notin P(c)$ satisfying $dist(p,p') < d_{cut}$,
where $p$ is a sampled point from $P(c)$ belong.
(We can deterministically decide $p$ in an arbitrary way.)

\vs
\noindent
\underline{\textbf{Local density computation.}}
S-Approx-DPC takes a similar approach to Ex-DPC.
For each cell of $G'$, S-Approx-DPC picks one point, say $p$, from $P(c)$, conducts a range search with query point $p$ and radius $d_{cut}$ on the $k$d-tree, and obtains $N(c)$.

From Lemma \ref{lemma_exdpc_local-density}, we have:

\begin{coro}	\label{corollray_sapproxdpc_local-density}
The time complexity of the local density computation in S-Approx-DPC is $O(\sum_{G'}(n^{1-1/d} + \rho_{avg}))$.
\end{coro}

\noindent
\underline{\textbf{Dependent point computation.}}
For each cell of $G'$, S-Approx-DPC has points that have not been picked during the local density computation.
S-Approx-DPC sets the picked point in the cell as their approximate dependent point.
As for piked points, we retrieve their approximate dependent points by a similar approach to that of Approx-DPC.
If there are some picked points $p_{i}$ that cannot find their approximate dependent points in the approach, we form temporary clusters.
We then find approximate dependent points while pruning temporary clusters that cannot have points with a small distance to $p_{i}$ and higher local density than $\rho_{i}$.
Below, we explain how to determine an approximate dependent point of a picked point.

\vs
\noindent
\textbullet $\,$ \underline{First phase.}
Let $P_{pick}$ be a set of picked points.
For the picked point $p_{i}$ in a cell $c$, if there exist some picked points in cells $\in N(c)$ with higher local density than $\rho_{i}$, an approximate dependent point of $p_{i}$ can be arbitrarily chosen from them.
In this case, its approximate dependent distance is bounded by $(1+\epsilon)d_{cut}$, from the definition of $G'$ and the fact that the distance between points in the same cell is at most $\epsilon \cdot d_{cut}$.
(This bound is useful for determining $\delta_{min}$.)

\vs
\noindent
\textbullet $\,$ \underline{Second phase.}
After the first phase, we have some picked points that do not have other picked points with higher local densities within $(1+\epsilon)d_{cut}$.
Let $P'_{pick}$ be a set of these points, and the remaining task is to retrieve an approximate dependent point of each point in $P'_{pick}$.
For each point $p_{i} \in P'_{pick}$, S-Approx-DPC computes $q'_{i}$, its nearest point with higher local density than $\rho_{i}$ among $P_{pick}$.

We here assume that $|P'_{pick}|^{2} \leq O(n)$.
(Otherwise, S-Approx-DPC employs the same approach as the exact dependent point computation in Approx-DPC.)
The detail is as follows:
\begin{enumerate}
    \setlength{\leftskip}{-2.0mm}
    \item	S-Approx-DPC forms \textit{temporary} clusters whose cluster centers are points in $P'_{pick}$, based on the dependency relationships computed in the first phase.
    \item	Let $T_{i}$ be a temporary cluster whose cluster center is $p_{i}$, that is, $T_{i}$ is a set of picked points that are reachable from $p_{i}$.
            S-Approx-DPC computes $r_{i} = \max_{p \in T_{i}} dist(p_{i},p)$.
    \item	S-Approx-DPC next computes $p' = \argmin dist(p_{i},p_{j})$, where $p_{j} \in P'_{pick}$ and $\rho_{i} < \rho_{j}$, for each $p_{i} \in P'_{pick}$.
    \item	Then, for $p_{i} \in P'_{pick}$, a temporary cluster $T_{j}$ cannot have $q'_{i}$ if $dist(p_{i},p_{j}) - r_{j} > dist(p_{i},p')$ or $\rho_{i} \geq \rho_{j}$, from triangle inequality and definition.
            Therefore, S-Approx-DPC retrieves $q'_{i}$ from the temporary clusters $T_{k}$ such that $dist(p_{i},p_{k}) - r_{k} \leq dist(p_{i},p')$ and $\rho_{i} < \rho_{k}$.
\end{enumerate}

This approach employs triangle inequality as with \cite{bai2017fast}, but utilizes it in a more effective way, because our approach exploits intermediate clusters in DPC but \cite{bai2017fast} uses $k$-means clustering, whose clusters are totally different from density-based clusters.
In addition, S-Approx-DPC has a theoretical efficiency:

\begin{lemm}	\label{lemma_sapproxdpc_dependent-point}
The time complexity of the dependent point computation in S-Approx-DPC is $O(\sqrt{n}|G'|)$ for an arbitrary fixed $d$, if $|P'_{pick}|^2 \leq O(n)$.
\end{lemm}

\noindent
\textsc{Proof.}
The first phase takes $O(|G'|)$ time for arbitrary fixed $d$, since $|N(c)| = O(1)$.
In the second phase, forming temporary clusters needs $O(|G'|)$ time.
Computing $r_{i}$ for each $T_{i}$ needs $O(\sum|T_{i}|) = O(|G'|)$ time, while we need $O(|P'_{pick}|^2) = O(n)$ time for computing $p'$ for each $p \in P'_{pick}$.
Last, computing $q'_{i}$ requires at most $O(|G'|)$ if it cannot prune any temporary clusters, thereby the worst case incurs $O(|P'_{pick}||G'|)$ time.
From $O(|P'_{pick}|) \leq O(\sqrt{n})$, the lemma holds.	\wsq

\vs
\noindent
\underline{\textbf{Analysis.}}
From Corollary \ref{corollray_sapproxdpc_local-density} and Lemma \ref{lemma_sapproxdpc_dependent-point}, we have:

\begin{theo}[\textsc{Time complexity of S-Approx-DPC}]
The time complexity of S-Approx-DPC is $O(\sum_{G'}(n^{1-1/d} + \rho_{avg}))$ for an arbitrary fixed $d$.
\end{theo}

We here discuss that the time complexity of S-Approx-DPC can be almost linear to $n$ for fixed parameters, if the distribution of $P$ does not change much and $|P'_{pick}|^2 \leq O(n)$.
This is the main property of S-Approx-DPC.
Fix $d_{cut}$, $\epsilon$, and $d$.
If the distribution of $P$ does not change, $|G'|$ can be considered as a constant.
This is because the distribution of cells in $G'$ simply follows the distribution of $P$, thereby new cells are rarely created, even if $n$ increases.
Then $O(\sum_{G'}(n^{1-1/d} + \rho_{avg})) \approx O(n^{1-1/d} + \rho_{avg})$.
The assumption that the distribution of $P$ does not change also yields that $\rho_{avg}$ grows linearly to $n$, resulting in the linear scalability of S-Approx-DPC to $n$ for fixed parameters.

Notice that, if $\epsilon$ becomes larger, $|G'|$ becomes smaller.
This leads to that the computation time is reduced but clustering accuracy may decrease.
Although S-Approx-DPC does not guarantee a trade-off relationship between efficiency and accuracy, S-Approx-DPC tends to have this relationship empirically.

Last, from essentially the same proof of Theorem \ref{theorem_approxdpc_space},

\begin{coro}[\textsc{Space complexity of S-Approx-DPC}]
The space complexity of S-Approx-DPC is $O(n)$ for an arbitrary fixed $d$.
\end{coro}

\noindent
\underline{\textbf{Implementation for parallel processing.}}
For local density computation, S-Approx-DPC takes the same approach as Ex-DPC.
Similarly to the local density computation, the operations in the dependent point computation of S-Approx-DPC consist of iterations, which can be simply parallelized by hash-partitioning or the same approach as the parallel local density computation.
That is, S-Approx-DPC is also parallel-friendly.

\section{Experiments}	\label{section_experiment}
Our experiments were conducted on a machine with dual 12-core Intel Xeon E5-2687W v4 processors (3.0GHz) that share a 512GB RAM.
By using hyper-threading, this machine can run at most 48 $(= 2 \times 12 \times 2)$ threads.
All evaluated algorithms were implemented in C++, and we used OpenMP for multi-threading.

\vs
\noindent
\underline{\textbf{Datasets.}}
We report the results on five synthetic dataset (\textit{Syn}, \textit{S1}, \textit{S2}, \textit{S3}, and \textit{S4}) and four real datasets (\textit{Airline}, \textit{Household}, \textit{PAMAP2}, and \textit{Sensor}).
The synthetic datasets are used for effectiveness evaluation.
Syn is a 2-dimensional dataset with 100,000 points, which was generated based on a random walk model introduced in \cite{gan2015dbscan}.
The domain of each dimension in Syn is $[0, 10^{5}]$.
Airline\footnote{\url{http://stat-computing.org/dataexpo/2009/}} is a 3-dimensional dataset with 5,810,462 points, and its domain of each dimension is $[0, 10^{6}]$.
Household and PAMAP2 are 4-dimensional datasets with respectively 2,049,280 and 3,850,505 points.
Last, Sensor is an 8-dimensional dataset with 928,991 points.
The domain of each dimension in Household, PAMAP2, and Sensor is the same as that of Syn, and they are from UCI machine learning archive\footnote{\url{https://archive.ics.uci.edu/ml/index.php}}.

\vs
\noindent
\underline{\textbf{Algorithms.}}
Our experiments evaluated the following algorithms:
\begin{itemize}
    \setlength{\leftskip}{-4.0mm}
    \item	\textit{Scan}: the straightforward algorithm introduced in \S \ref{section_problem-definition}.
    \item	\textit{R-tree $+$ Scan}: a variant of Scan which computes local density by using an in-memory R-tree.
    \item	\textit{LSH-DDP} \cite{zhang2016efficient}: a state-of-the-art approximation algorithm.
    \item	\textit{CFSFDP-A} \cite{bai2017fast}: a state-of-the-art exact algorithm.
    \item	\textit{Ex-DPC}: our exact algorithm introduced in \S \ref{section_exdpc}.
    \item	\textit{Approx-DPC}: our approximation algorithm introduced in \S \ref{section_approxdpc}.
    \item	\textit{S-Approx-DPC}: our approximation algorithm introduced in \S \ref{section_sapproxdpc}.
\end{itemize}
Codes are available in a GitHub repository\footnote{\url{https://github.com/amgt-d1/DPC}}.
We followed the original paper to set the inner parameters of LSH-DDP and CFSFDP-A.
Table \ref{table_time-complexity} shows that the dependent distance computation of CFSFDP-A is slower than that of Scan.
We hence used the approach of Scan for computing dependent distances in CFSFDP-A.

Other approximation algorithms, FastDPeak \cite{chen2020fast}, DPCG \cite{xu2018dpcg}, and CFSFDP-DE \cite{bai2017fast} were also tested.
We confirmed that FastDPeak and DPCG are slow (e.g., FastDPeak and DPCG respectively took 8114 and 14390 seconds on Airline at the default parameter setting) and significantly outperformed by our \textit{exact} algorithm.
In addition, we observed that the clustering accuracy (Rand index) of CFSFDP-DE is quite low (e.g., 0.18 on PAMAP2).
We therefore omit the detailed results of these algorithms to keep space limitation\footnote{As existing works, e.g., \cite{zhang2016efficient}, show that the output of DPC is different from those of the other clustering algorithms, we do not test them.}.

\subsection{Effectiveness Evaluation}
\begin{figure}[!t]
	\begin{center}
	\subfigure[Original]{%
		\includegraphics[width=0.485\linewidth]{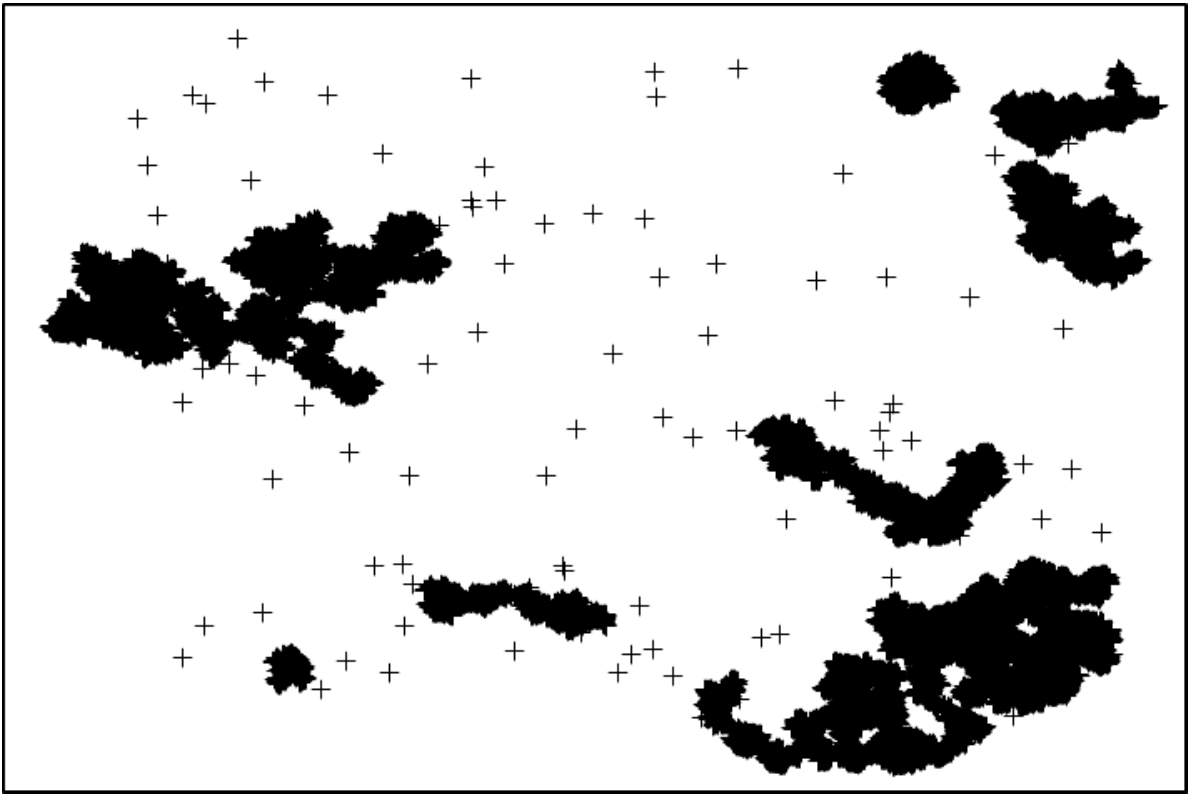}	\label{fig_syn_original}}
        \subfigure[Ex-DPC]{%
		\includegraphics[width=0.485\linewidth]{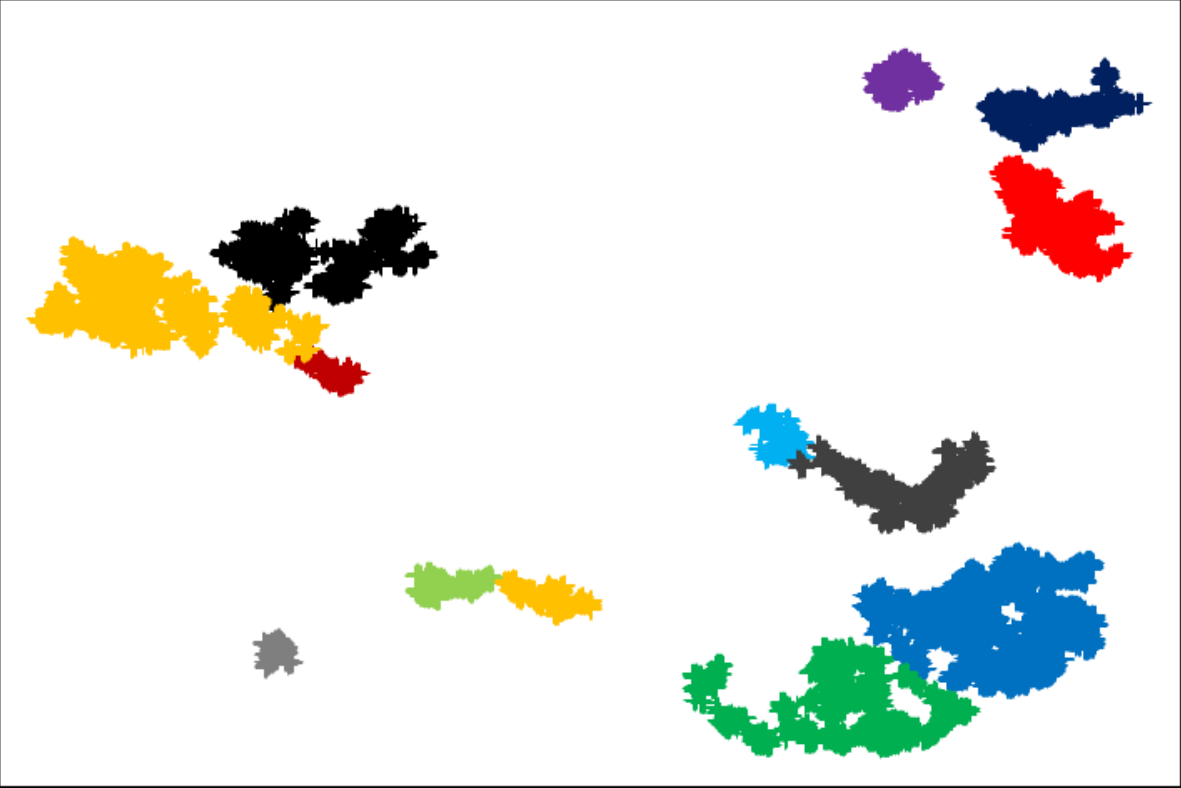}		\label{fig_syn_exact}}
        \subfigure[LSH-DDP]{%
		\includegraphics[width=0.485\linewidth]{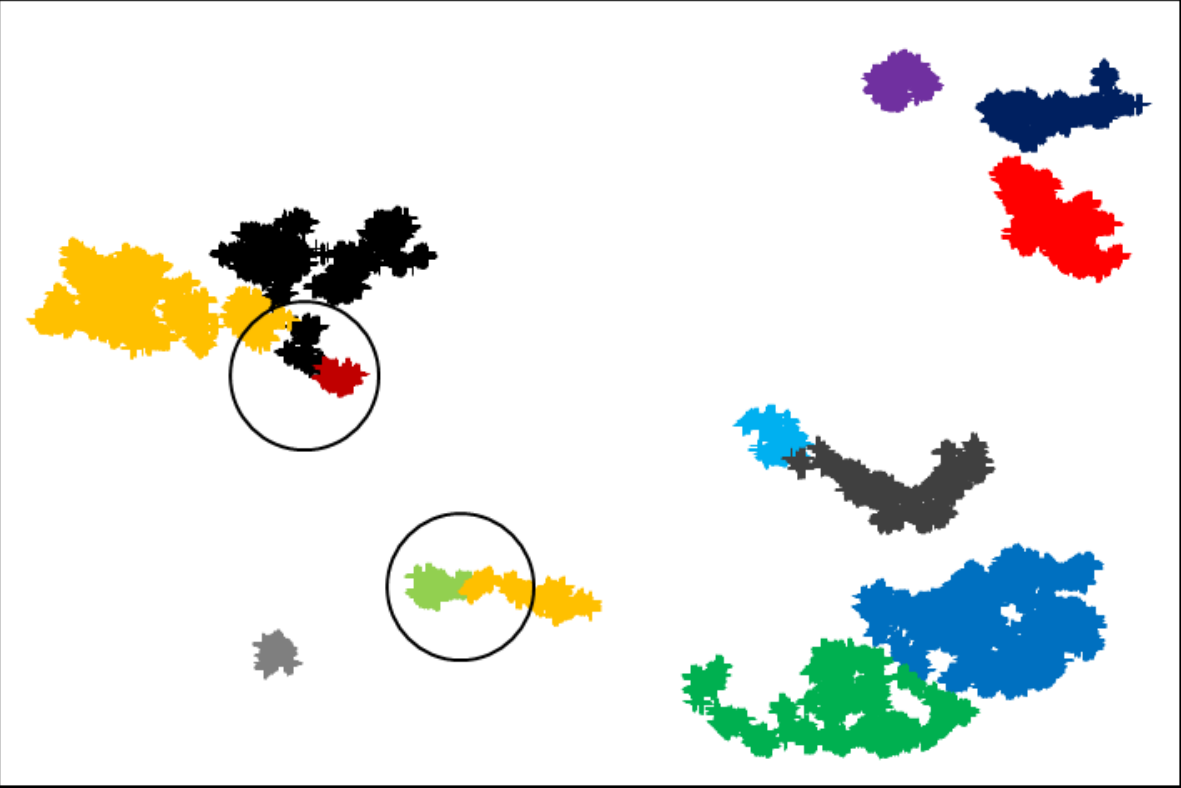}		\label{fig_syn_lshddp}}
        \subfigure[Approx-DPC]{%
		\includegraphics[width=0.485\linewidth]{Figure/syn_exact.pdf}		\label{fig_syn_approxdpc}}
        \subfigure[S-Approx-DPC ($\epsilon = 0.2$)]{%
		\includegraphics[width=0.485\linewidth]{Figure/syn_exact.pdf}		\label{fig_syn_sapproxdpc}}
        \subfigure[S-Approx-DPC ($\epsilon = 1.0$)]{%
		\includegraphics[width=0.485\linewidth]{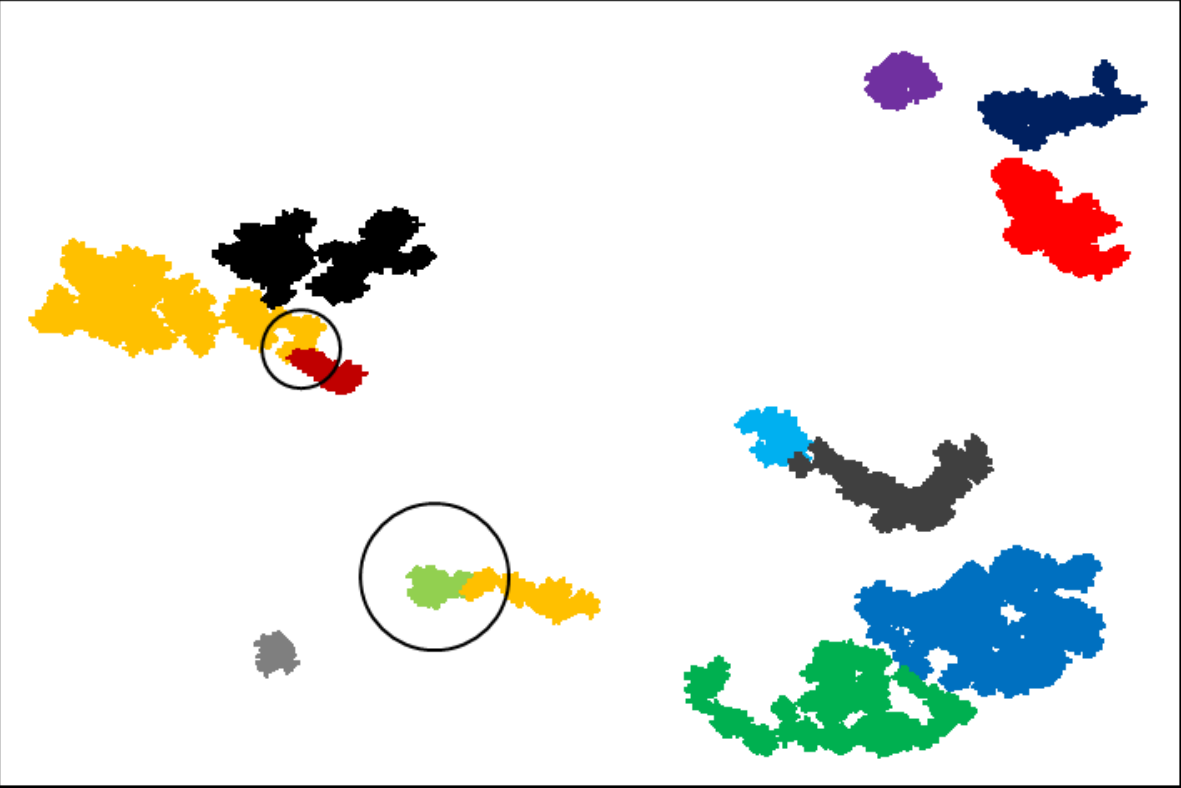}	\label{fig_syn_sapproxdpc_}}
        \caption{2D visualization of the clustering result of each algorithm ($d_{cut} = 250$)}	\label{figure_visual}
	\end{center}
\end{figure}

\textbf{2D visualization.}
To visually understand the effectiveness of our approximation algorithms, we used Syn, which is depicted in Figure \ref{fig_syn_original} and has 13 density-peaks.
We set $d_{cut} = 250$.
In addition, $\rho_{min}$ is specified so that noises can be clearly removed, while $\delta_{min}$ is specified so that we have 13 clusters.
The clustering result of Ex-DPC, which is shown in Figure \ref{fig_syn_exact} where points in each cluster are illustrated by the same color, is used as ground-truth.

Let us first focus on Figure \ref{fig_syn_lshddp}, the clustering result of the state-of-the-art approximation algorithm LSH-DDP.
It has two major differences from that of Ex-DPC (specified by circles).
Since LSH-DDP approximates both local density and dependent point, for a point $p_{i}$, it may decide that an approximate dependent point of $p_{i}$ is $p_{j}$ even if we indeed have $\rho_{i} \geq \rho_{j}$.
This makes it hard for analysts to know why a cluster (e.g., the black one in LSH-DDP) is different from the exact one.
This is a drawback of LSH-DDP.

We look at Figure \ref{fig_syn_approxdpc}, the clustering result of Approx-DPC.
Actually this is the same clustering result as that of Ex-DPC.
This result demonstrates that our key idea (for $p \in P$, compute an approximate dependent point if there exists a close point with higher local density; otherwise, compute the exact one) is promising.

We turn our attention to the clustering result of S-Approx-DPC.
Figures \ref{fig_syn_sapproxdpc} and \ref{fig_syn_sapproxdpc_} respectively illustrate the cases where $\epsilon = 0.2$ and $\epsilon = 1.0$.
The case where $\epsilon = 0.2$ also returns the correct clustering result.
This is reasonable, because $\epsilon = 0.2$ creates many cells and each approximate dependent point tends to belong to the same cluster as the exact dependent point.
On the other hand, the case where $\epsilon = 1.0$ has one major difference and one minor difference.
If a point has an approximate dependent point, it may belong to a different cluster from the exact one.
In S-Approx-DPC, the approximate dependent distance of each picked point is guaranteed to be larger than (or equal to) the exact dependent distance (recall that picked points have \textit{exact} local densities).
For a point $p$ that exists at a border between different clusters, even a small distance difference influences its cluster label.
For instance, the nearest neighbor point with higher local density than $\rho$ belongs to cluster $C_{1}$, but a point close to $p$ belongs to $C_{2}$.
This observation derives the difference specified by the circles in Figure \ref{fig_syn_sapproxdpc_}.

\vs
\noindent
\underline{\textbf{Quantitative evaluation.}}
We next investigate the accuracy of the approximation algorithms.
We used Rand index to measure the accuracy of each approximation algorithm under the same parameter setting as Ex-DPC (i.e., the clustering result of Ex-DPC is the ground truth).

First, we investigate the robustness of the approximation algorithms to noise rate.
We varied the noise rate of Syn, and Table \ref{table_rand-index_noise} shows the result, where $\epsilon = 1.0$ for S-Approx-DPC.
(Bold shows the winner.)
From the result, we see that their accuracy is still high even when Syn has many noises (e.g., the rate is 0.16), showing the robustness to noises.

\begin{table}[!t]
\begin{center}
	\caption{Rand index of LSH-DDP, Approx-DPC, and S-Approx-DPC on Syn with different noise rate}	\label{table_rand-index_noise}
	\begin{tabular}{c||c|c|c} \hline
        Noise rate  & LSH-DDP	& Approx-DPC	    & S-Approx-DPC	    \\ \hline \hline
        0.01		& 0.999 	& \textbf{1.000}    & 0.995		        \\ \hline
        0.02	    & 0.980		& \textbf{0.984}    & 0.980		        \\ \hline
        0.04	    & 0.979		& \textbf{0.983}    & \textbf{0.983}    \\ \hline
        0.08	    & 0.981		& \textbf{0.982}    & \textbf{0.982}    \\ \hline
        0.16	    & 0.969		& \textbf{0.976}    & 0.970		        \\ \hline
	\end{tabular}
\end{center}
\end{table}
\begin{table}[!t]
\begin{center}
	\caption{Rand index of LSH-DDP, Approx-DPC, and S-Approx-DPC on S1, S2, S3, and S4}	\label{table_rand-index_overlap}
	\begin{tabular}{c||c|c|c} \hline
                Dataset  & LSH-DDP	& Approx-DPC	    & S-Approx-DPC	\\ \hline \hline
        		S1		& 0.996 	& \textbf{1.000}    & 0.999		    \\ \hline
                S2	    & 0.994		& \textbf{0.998}    & 0.996		    \\ \hline
                S3	    & 0.989		& \textbf{0.999}    & 0.988		    \\ \hline
                S4	    & 0.979		& \textbf{0.990}    & 0.981		    \\ \hline
	\end{tabular}
\end{center}
\end{table}
\begin{table}[!t]
\begin{center}
	\caption{Rand index of LSH-DDP and Approx-DPC on real datasets}	\label{table_rand-index_approxdpc}
	\begin{tabular}{c||c|c|c|c} \hline
                			& Airline	        & Household	        & PAMAP2	        & Sensor	        \\ \hline \hline
        		LSH-DDP		& 0.938		        & 0.983		        & 0.951		        & 0.902		        \\ \hline
                Approx-DPC	& \textbf{0.999}	& \textbf{0.996}	& \textbf{0.996}	& \textbf{0.960}    \\ \hline
	\end{tabular}
\end{center}
\end{table}
\begin{table}[!t]
\begin{center}
	\caption{Running time [sec] vs. accuracy (Rand index) of S-Approx-DPC}	\label{table_tradeoff}
	\begin{tabular}{c||c|c|c|c} \hline
 				& \multicolumn{2}{|c|}{Airline}	& \multicolumn{2}{|c}{Household}	\\ \hline \hline
                $\epsilon$	& Time		& Rand index	& Time		& Rand index	\\ \hline
                0.2	    	& 32.178	& 0.998		    & 59.597	& 0.995			\\ \hline
                0.4		    & 29.992	& 0.996		    & 27.637	& 0.994			\\ \hline
                0.6		    & 25.935	& 0.985		    & 16.470	& 0.994			\\ \hline
                0.8		    & 20.401	& 0.976		    & 11.097	& 0.993			\\ \hline
                1.0	    	& 16.449	& 0.969		    & 7.527		& 0.991			\\ \hline
	\end{tabular}
\end{center}
\end{table}

\begin{figure*}[!t]
	\begin{center}
    \includegraphics[width=0.80\linewidth]{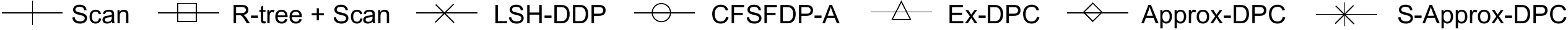}
    \vspace{-1.0mm}
    
		\subfigure[Airline]{%
		\includegraphics[width=0.240\linewidth]{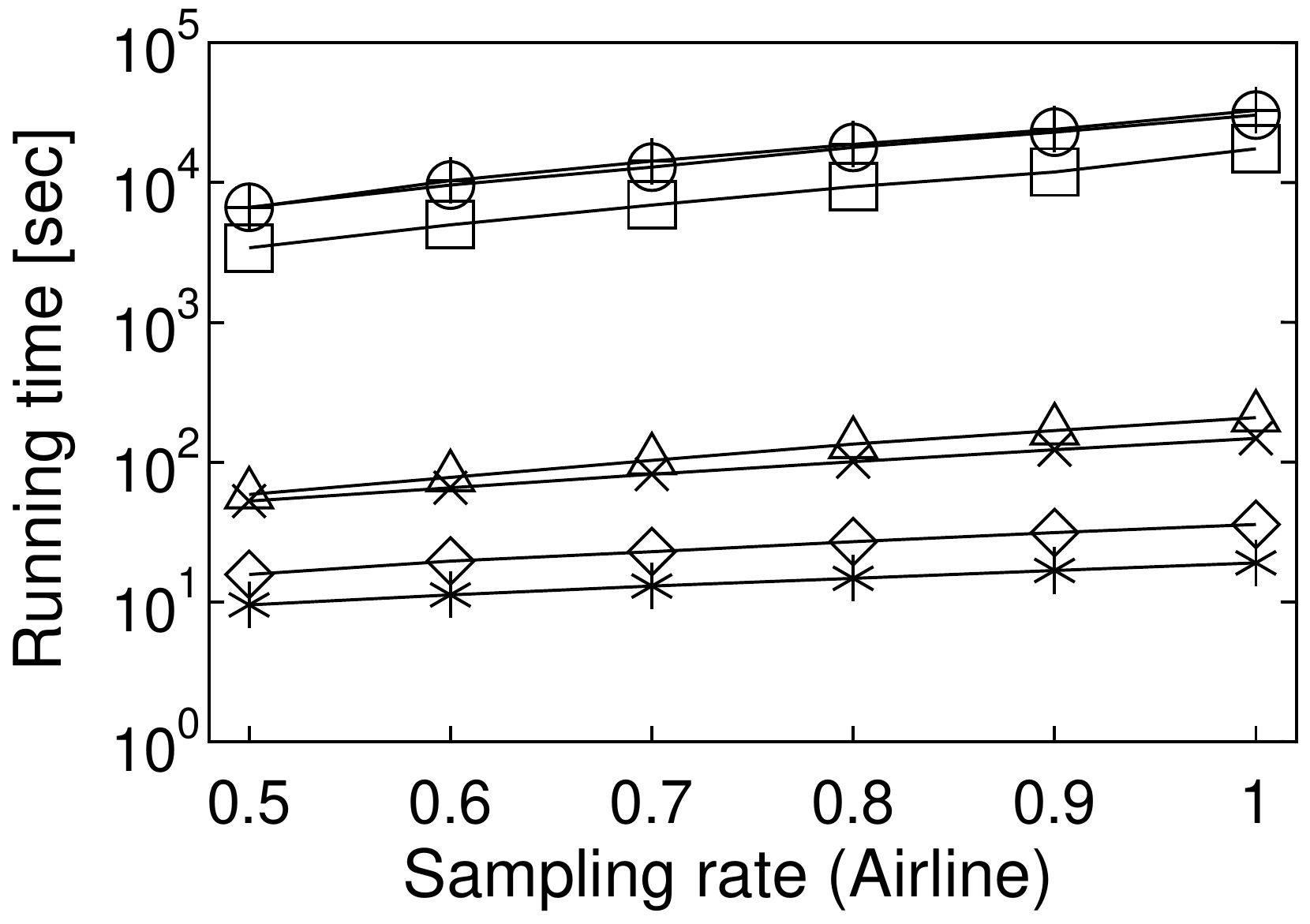}	\label{fig_sampling_airline}}
        \subfigure[Household]{%
		\includegraphics[width=0.240\linewidth]{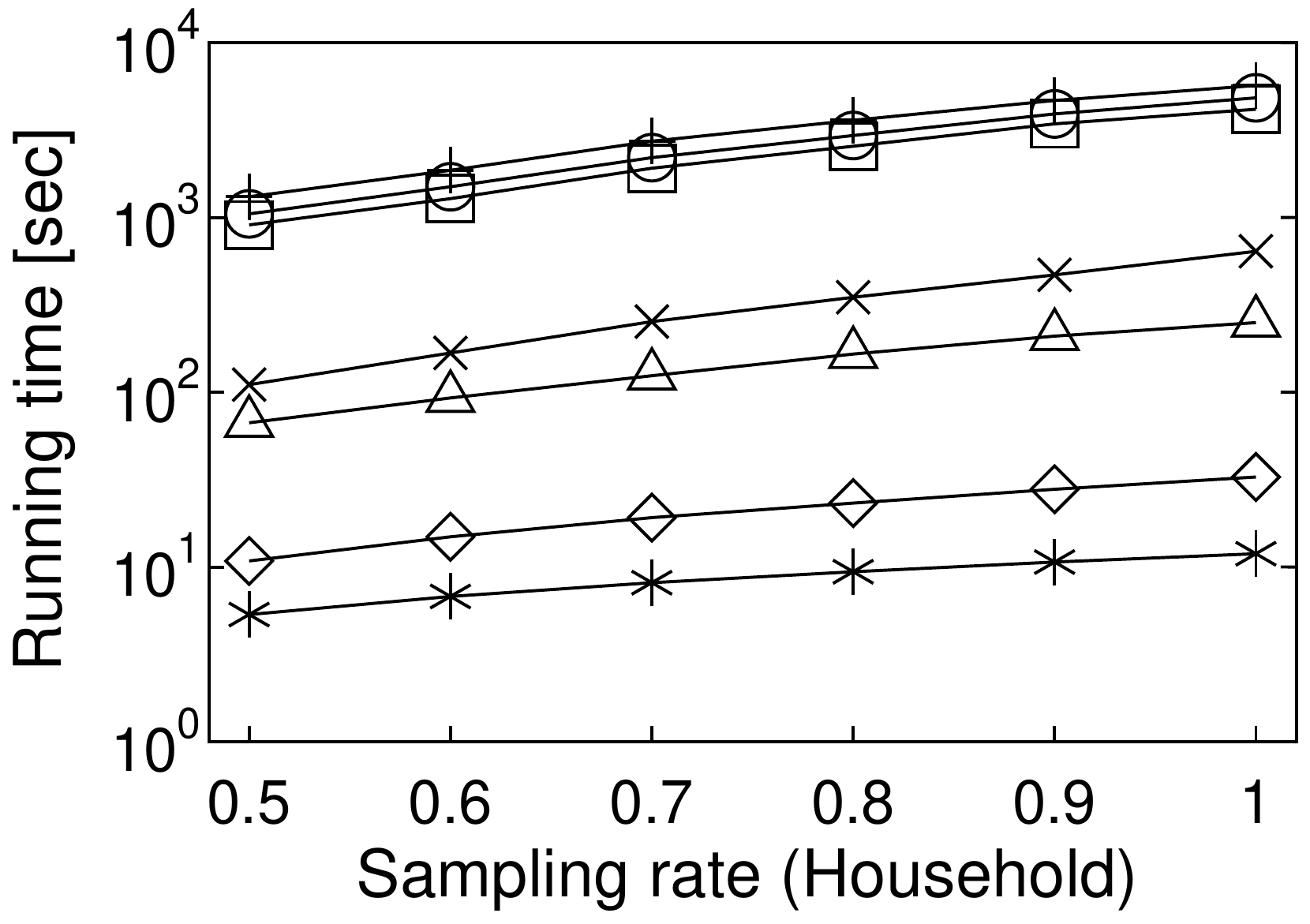}	\label{fig_sampling_household}}
        \subfigure[PAMAP2]{%
		\includegraphics[width=0.240\linewidth]{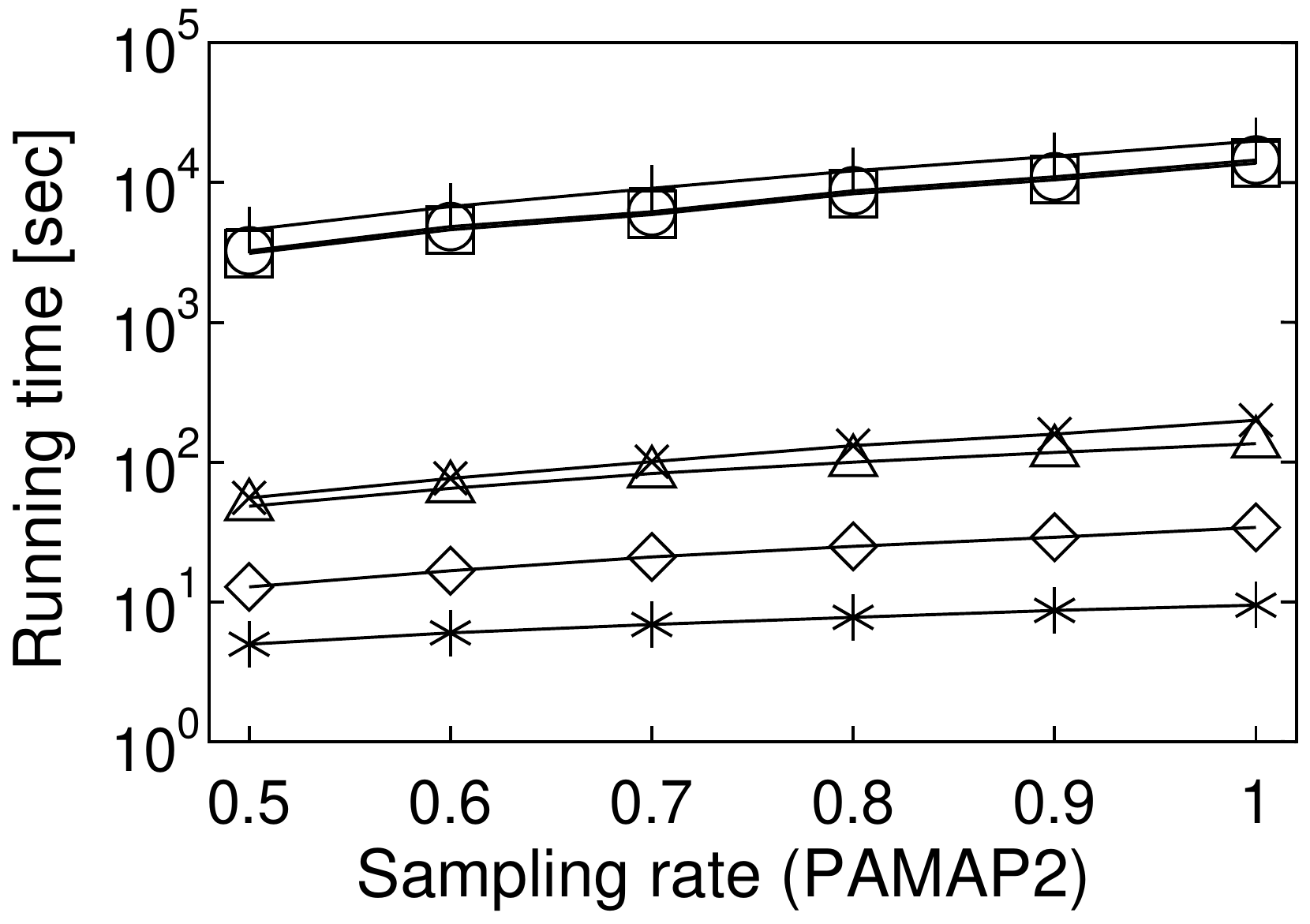}		\label{fig_sampling_pamap2}}
        \subfigure[Sensor]{%
		\includegraphics[width=0.240\linewidth]{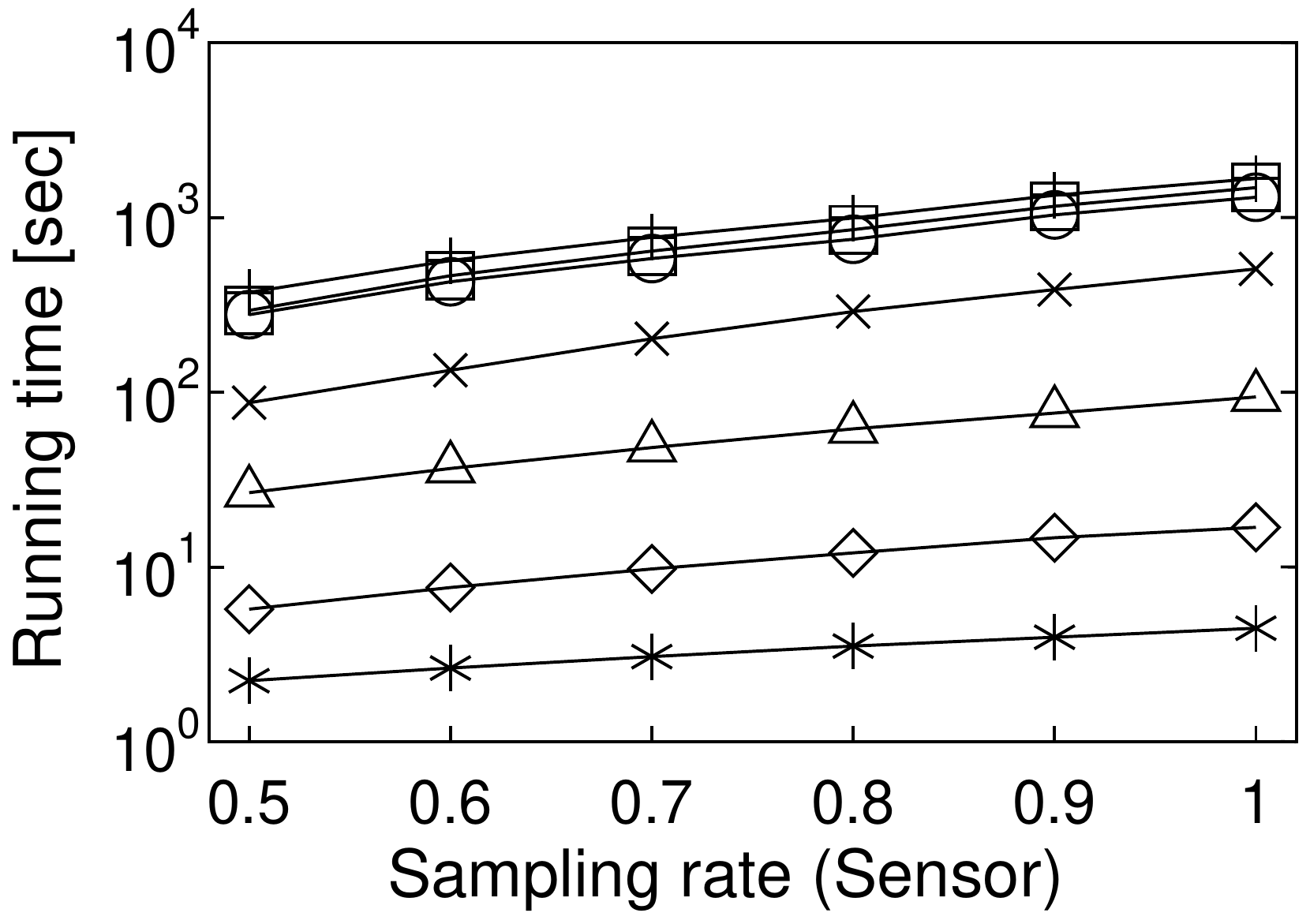}		\label{fig_sampling_seensor}}
        \caption{Impact of cardinality (sampling rate)}
        \label{figure_sample}
	\end{center}
\end{figure*}
\begin{table*}[!t]
\begin{center}
    \caption{Decomposed time [sec] (parameters are default ones)}
    \label{table_decomposed-time}
	\begin{tabular}{c||c|c|c|c|c|c|c|c} \hline
 				                & \multicolumn{2}{|c|}{Airline}	        & \multicolumn{2}{|c|}{Household}	& \multicolumn{2}{|c|}{PAMAP2}		& \multicolumn{2}{|c}{Sensor}		\\ \hline \hline
                Algorithm		& $\rho$ comp.	    & $\delta$ comp.	& $\rho$ comp.	& $\delta$ comp.	& $\rho$ comp.	& $\delta$ comp.	& $\rho$ comp.	& $\delta$ comp.	\\ \hline
                Scan			& 15492.70		    & 17310.40			& 1703.37		& 3989.77			& 6114.82		& 13717.60			& 492.60		& 1178.27			\\ \hline
                R-tree + Scan	& 128.28		    & -					& 174.82		& -					& 36.76			& -					& 304.10		& -					\\ \hline
                LSH-DDP			& 90.54			    & 56.86				& 225.69		& 414.37			& 99.67			& 98.38				& 148.84		& 358.84			\\ \hline
                CFSFDP-A		& 13091.20		    & -					& 850.34		& -					& 776.94		& -					& 127.89		& -					\\ \hline
                Ex-DPC			& 79.20			    & 129.56			& 67.27			& 182.47			& 36.68			& 97.45				& 89.93			& 5.40				\\ \hline
                Approx-DPC		& 25.09			    & 3.77				& 22.12			& 8.66				& 18.88			& 12.29				& 14.72			& 1.58				\\ \hline
                S-Approx-DPC	& \textbf{11.24}	& \textbf{1.16}		& \textbf{7.92}	& \textbf{0.74}		& \textbf{6.08}	& \textbf{0.72}		& \textbf{3.56}	& \textbf{0.27}		\\ \hline
	\end{tabular}
\end{center}
\end{table*}

Furthermore, we investigate the robustness to the degree of cluster overlapping by using S1, S2, S3, and S4.
These datasets have 15 Gaussian clusters and the same cardinality, whereas the degree of cluster overlapping of Sx increases as x increases, see \cite{franti2018k}.
That is, this experiment studies the impact of cluster overlapping rate.
Table \ref{table_rand-index_overlap} shows that they provide almost perfect results even when the degree is high, i.e., on S4 (we confirmed that Ex-DPC and the approximation algorithms provide 15 clusters).
This result demonstrates the robustness of our algorithms to cluster overlapping.
(We confirmed that the noise rate and cluster overlapping degree do not affect the efficiencies of our algorithms.)

Table \ref{table_rand-index_approxdpc} shows the Rand index of LSH-DDP and Approx-DPC on real datasets.
We set $d_{cut} = 1000$ for Airline, Household, and PAMAP2, and $d_{cut} = 5000$ for Sensor (these are default values of $d_{cut}$).
For each dataset, we specified $\rho_{min}$ and $\delta_{min}$ based on the discussion in Section \ref{section_preliminary}.
As with the 2D visualization, Approx-DPC provides a highly accurate clustering result and beats LSH-DDP.

We study the impact of $\epsilon$ on the Rand index of S-Approx-DPC, and the result is shown in Table \ref{table_tradeoff} (we omit the results on PAMAP2 and Sensor, because they are similar to those on Airline and Household).
The Rand index decreases with increase of $\epsilon$.
Because the cell size of $G'$ becomes larger as $\epsilon$ increases, the approximate dependent point of each point can be more rough.
As mentioned before, the approximate dependent distance of each picked point is larger than its exact dependent distance.
For the same $\delta_{min}$ as Ex-DPC, therefore, S-Approx-DPC may provide more clusters, which also degrades the clustering accuracy.
However, the impact of $\epsilon$ is small, i.e., the decrease in the Rand index is slight, and S-Approx-DPC beats LSH-DDP as well as Approx-DPC but its Rand index does not reach to that of Approx-DPC.
Table \ref{table_tradeoff} also shows the running time of S-Approx-DPC with 12 threads.
From the relationship between running time and Rand index, we used 0.8, 0.8, 0.8, and 0.6 as $\epsilon$ for Airline, Household, PAMAP2, and Sensor, respectively.

\subsection{Efficiency Evaluation}	\label{section_efficiency-evaluation}
We report the running time of each algorithm.
The default number of threads is 12.

\begin{figure*}[!t]
	\begin{center}
	    \includegraphics[width=0.675\linewidth]{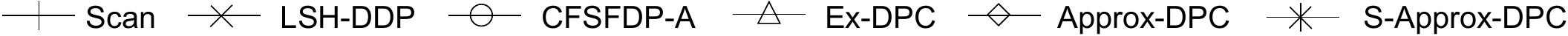}
        \vspace{-1.0mm}
        
		\subfigure[Airline]{%
		\includegraphics[width=0.240\linewidth]{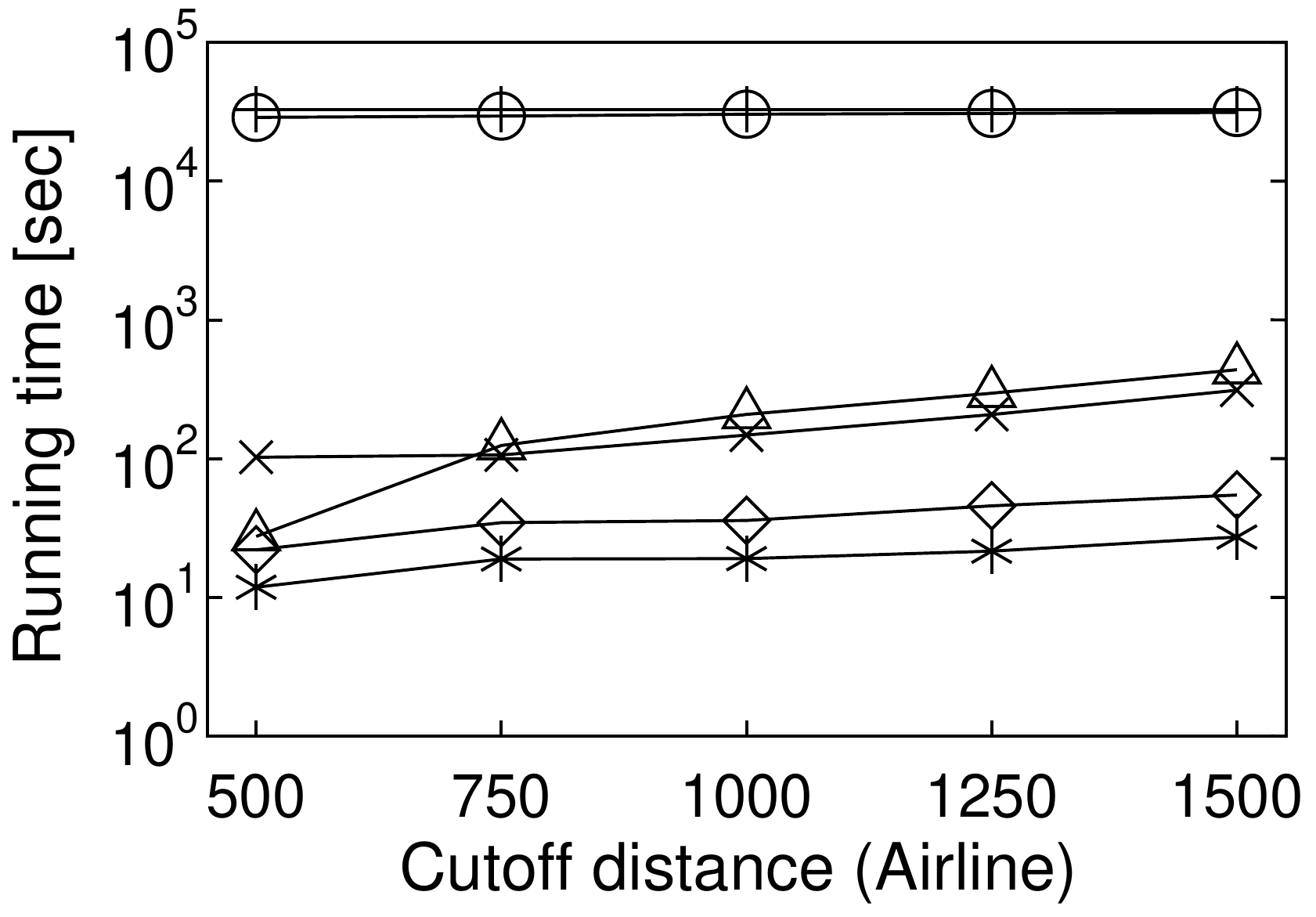}	\label{fig_cutoff_airline}}
        \subfigure[Household]{%
		\includegraphics[width=0.240\linewidth]{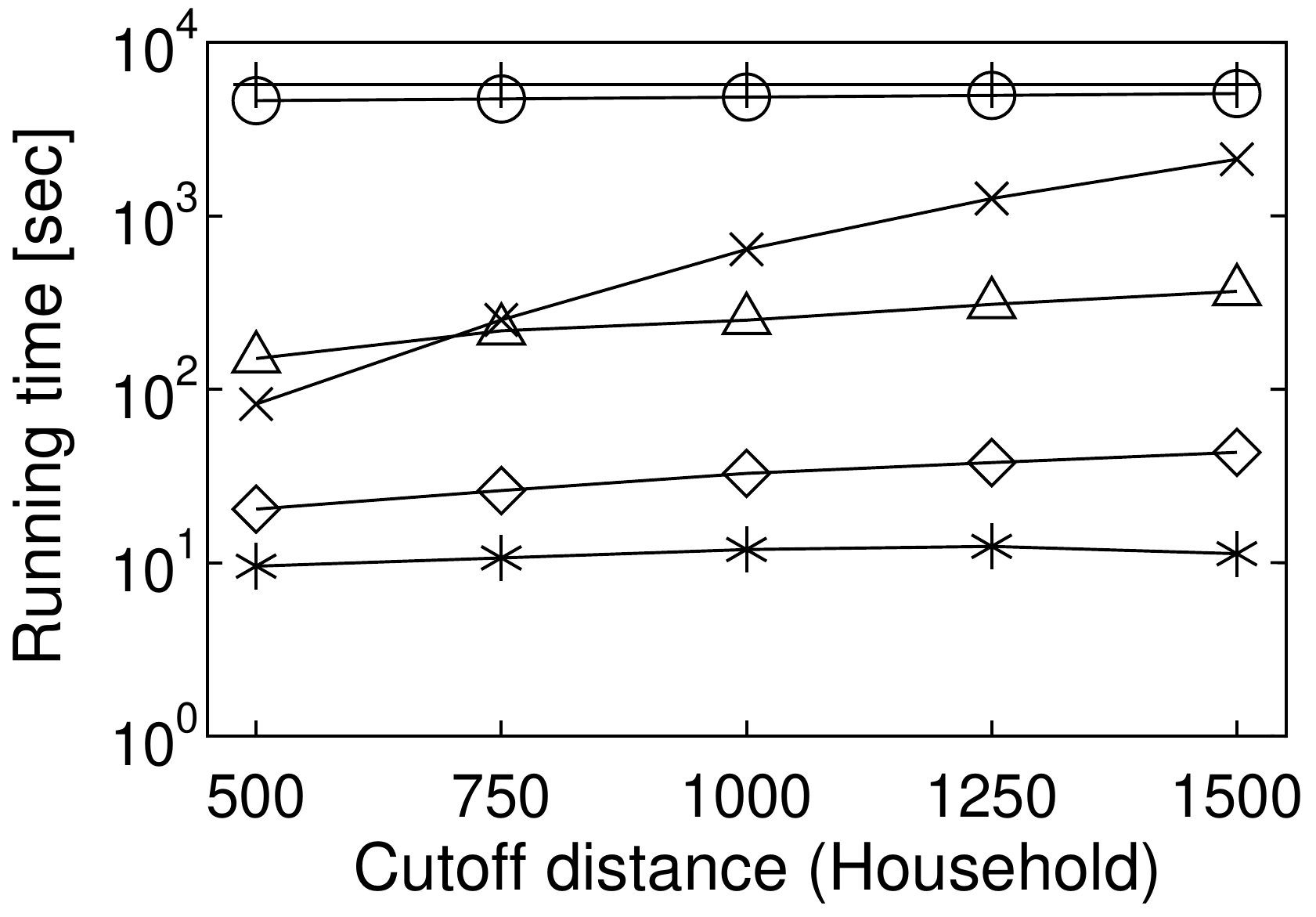}	\label{fig_cutoff_household}}
        \subfigure[PAMAP2]{%
		\includegraphics[width=0.240\linewidth]{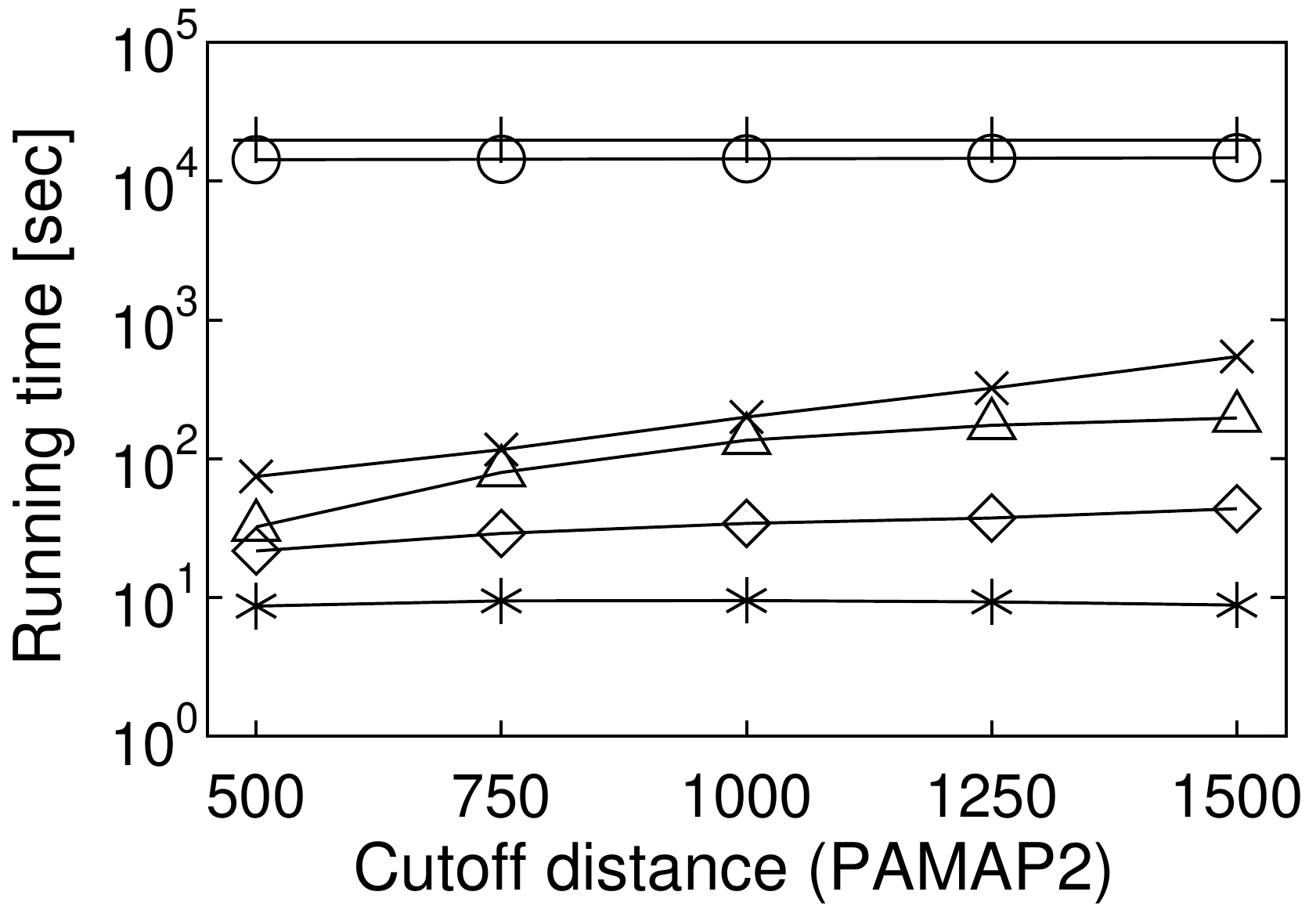}	\label{fig_cutoff_pamap2}}
        \subfigure[Sensor]{%
		\includegraphics[width=0.240\linewidth]{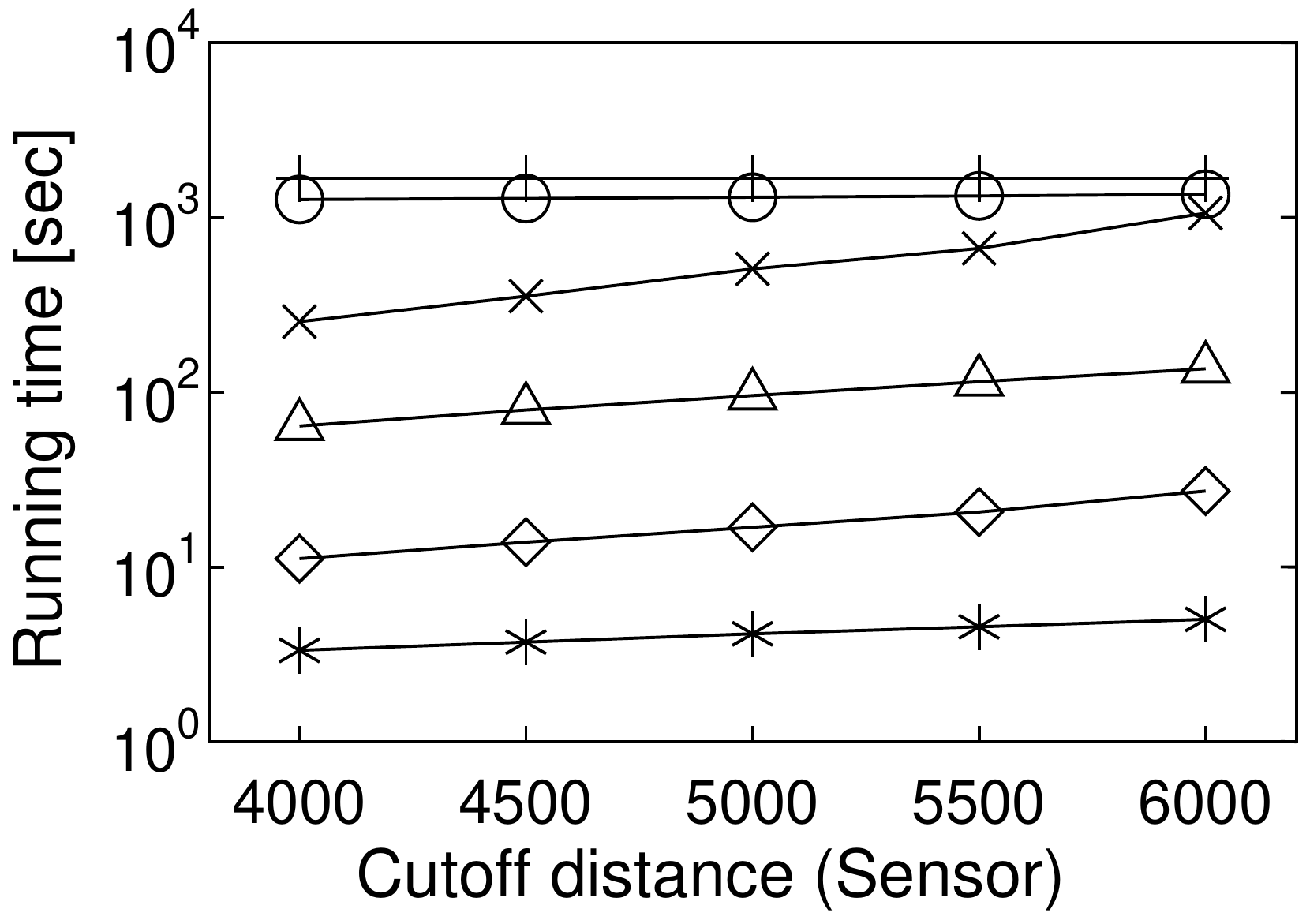}	\label{fig_cutoff_seensor}}
        \caption{Impact of $d_{cut}$}	\label{figure_cutoff}
	\end{center}
\end{figure*}
\begin{figure*}[!t]
	\begin{center}
		\subfigure[Airline]{%
		\includegraphics[width=0.240\linewidth]{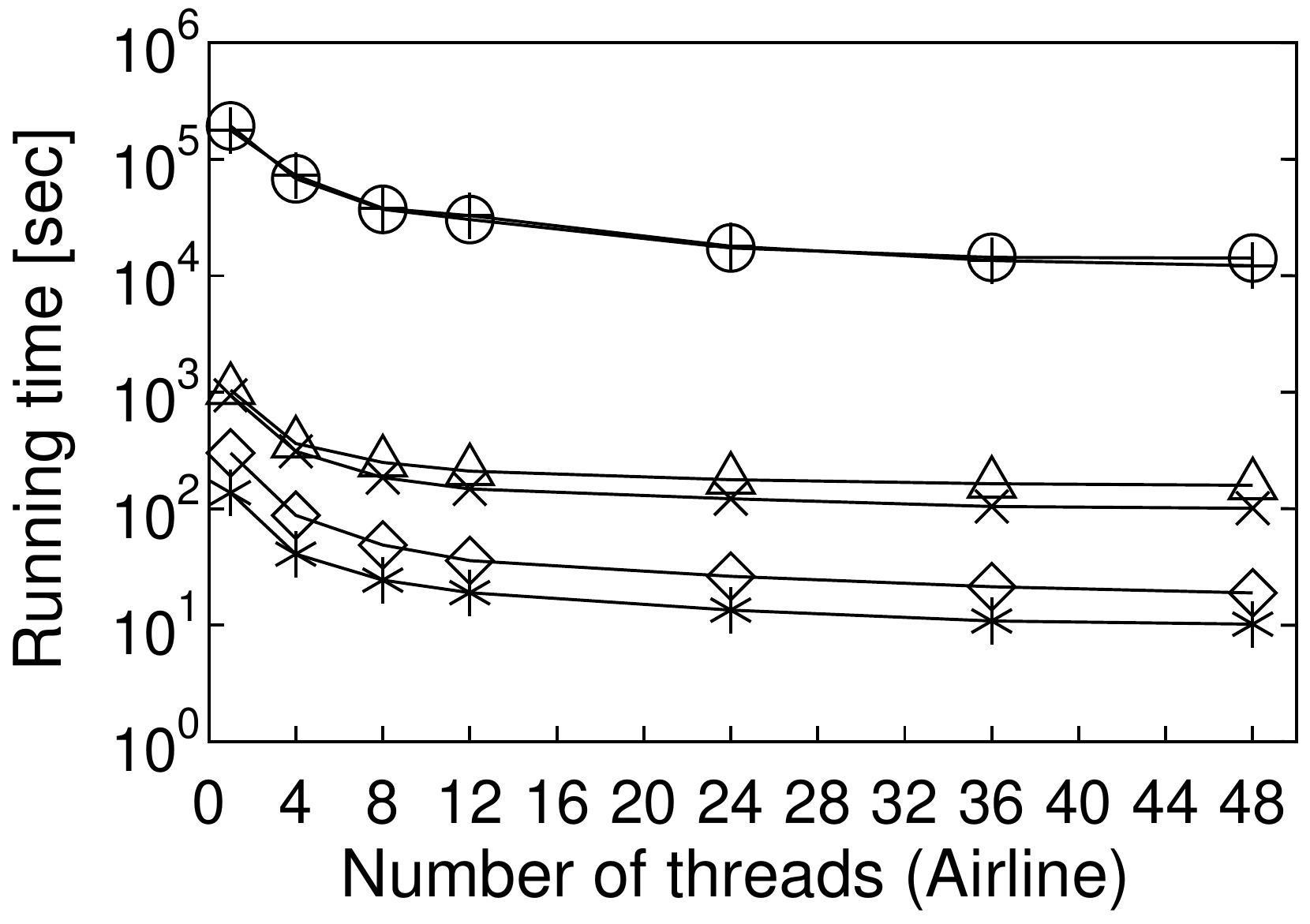}	\label{fig_thread_airline}}
        \subfigure[Household]{%
		\includegraphics[width=0.240\linewidth]{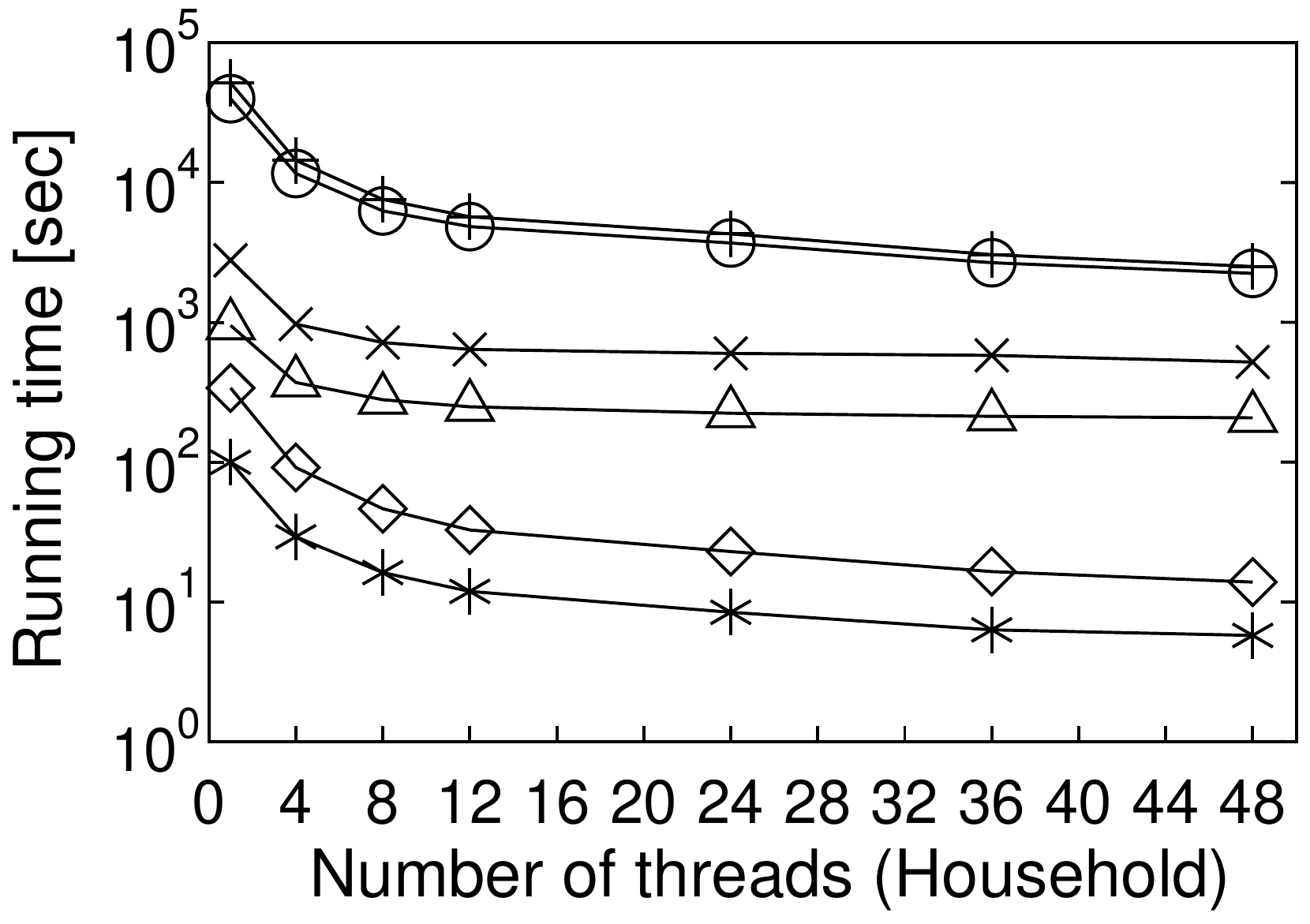}	\label{fig_thread_household}}
        \subfigure[PAMAP2]{%
		\includegraphics[width=0.240\linewidth]{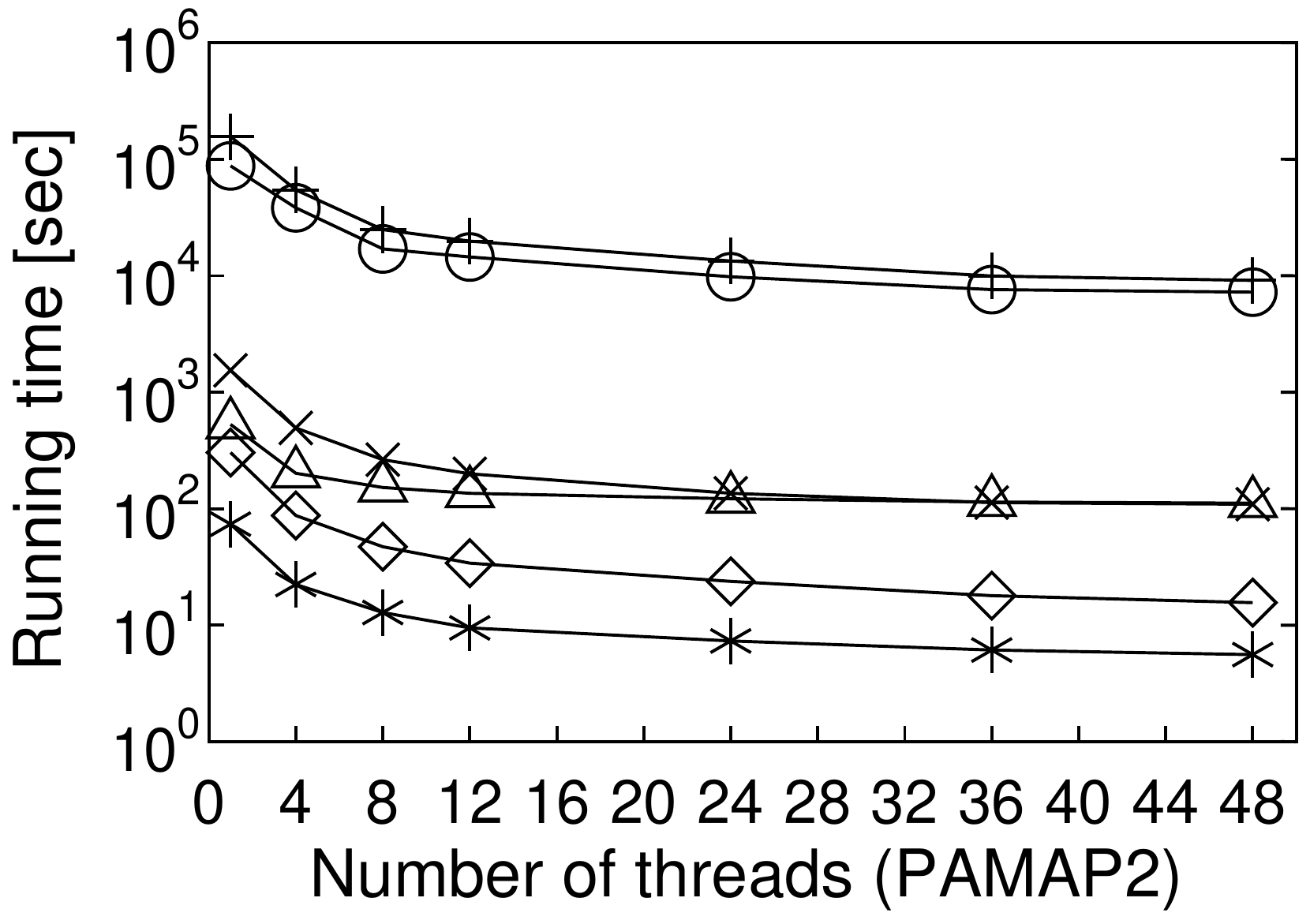}	\label{fig_thread_pamap2}}
        \subfigure[Sensor]{%
		\includegraphics[width=0.240\linewidth]{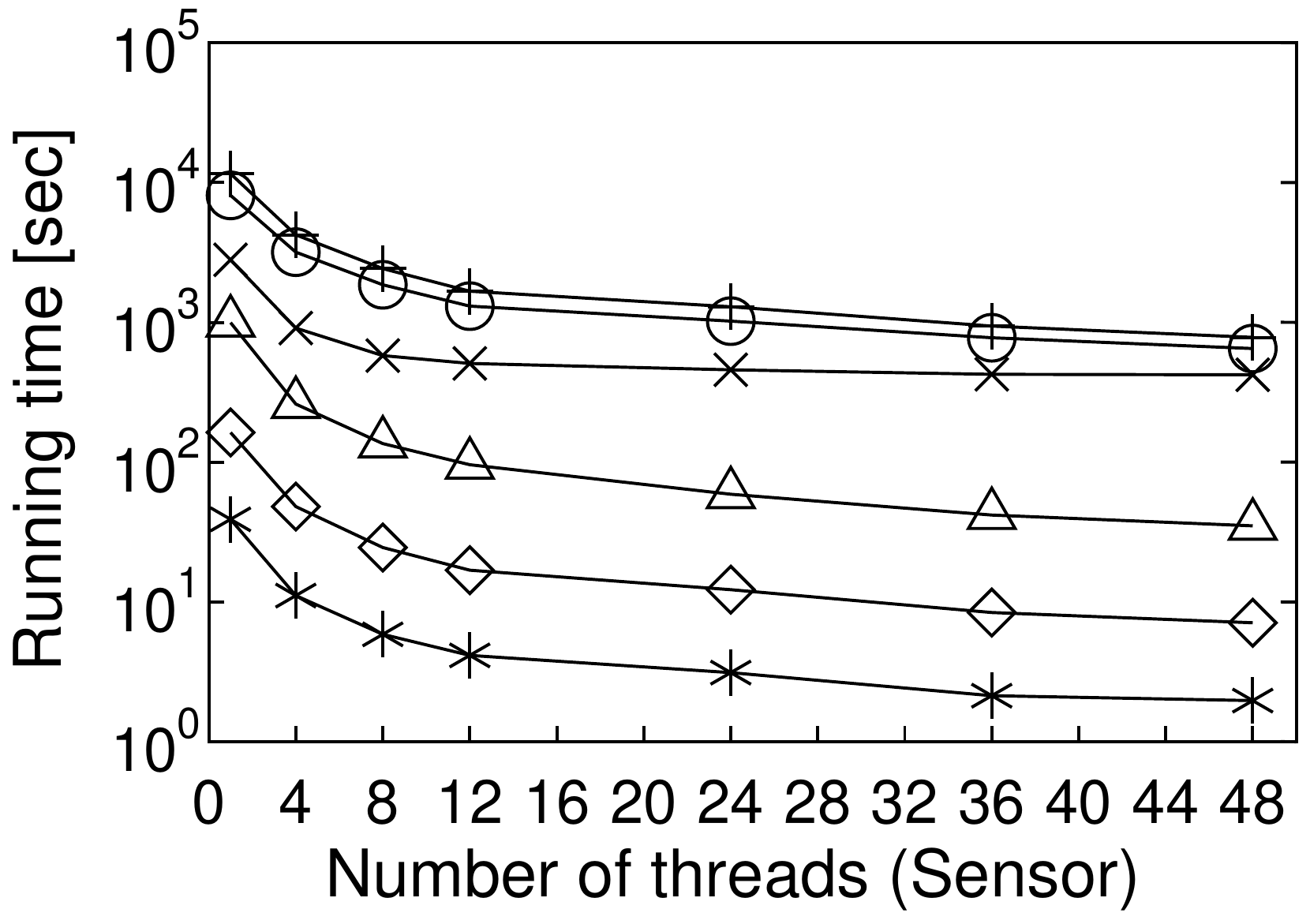}	\label{fig_thread_seensor}}
        \caption{Impact of number of threads}	\label{figure_thread}
	\end{center}
\end{figure*}

\vs
\noindent
\underline{\textbf{Impact of cardinality.}}
We first investigate the scalability of each algorithm to the number of points in a dataset.
We varied the number of points in each dataset via uniform sampling, i.e., by varying sampling rate (the other parameter are fixed by their default values).
Figure \ref{figure_sample} plots the result.

Let us focus on the exact algorithms (Scan, R-tree + Scan, CFSFDP-A, and Ex-DPC).
Ex-DPC significantly outperforms the other exact algorithms.
For example, when sampling rate is 1, the running time of Ex-DPC is 145.9, 19.3, 106.8, and 13.7 times faster than that of CFSFDP-A on Airline, Household, PAMAP2, and Sensor, respectively.
Table \ref{table_decomposed-time} exhibits the time to compute local densities and dependent points of all points at the default parameter setting.
(Since R-tree + Scan and CFSFDP-A employ the same dependent point computation as Scan, their $\delta$ comp. are blank but the same as that of Scan.)
We see that Ex-DPC clearly improves both the computation, compared with the other exact algorithms, although Ex-DPC employs simple approaches.
The R-tree alleviates the cost of local density computation, but R-tree + Scan still suffers from the quadratic cost of dependent point computation.
CFSFDP-A also suffers from the quadratic time computation, and its running time (i.e., local density computation time) has a small gain against Scan, because its filtering technique is sensitive to noises.
On the other hand, Ex-DPC scales much better, as it is sub-quadratic to $n$.
Interestingly, Ex-DPC beats the existing approximation algorithm LSH-DDP on Household, PAMAP2, and Sensor.
(We hereafter omit the result of R-tree + Scan, as its behavior is similar to that of Scan.)

Consider approximation algorithms.
Approx-DPC outperforms the exact algorithms and LSH-DDP.
When the sampling rate is 1, the running time of Approx-DPC is 4.1, 19.6, 5.8, and 30.1 times faster than that of LSH-DDP on Airline, Household, PAMAP2, and Sensor, respectively.
Table \ref{table_decomposed-time} demonstrates that our joint range search improves the iterative range search (see $\rho$ comp. of Ex-DPC and Approx-DPC).
Besides, Approx-DPC scales better than LSH-DDP.
S-Approx-DPC further improves the efficiency of DPC and actually scales linearly to sampling rate, as analyzed in \S \ref{section_sapproxdpc}.

\vs
\noindent
\underline{\textbf{Impact of $d_{cut}$.}}
The main parameter of DPC is $d_{cut}$, thus we next study its impact.
Figure \ref{figure_cutoff} depicts the experimental result.
The first observation is that Scan and CFSFDP-A are not sensitive to $d_{cut}$.
This is reasonable, because they compute local density and dependent point based on scanning $P$.
Second, LSH-DDP is very sensitive to $d_{cut}$.
This parameter affects the bucket size of locality-sensitive hashing, and some buckets have many points when $d_{cut}$ is large.
This observation can be understood from Table \ref{table_time-complexity}.
Our algorithms are also affected by $d_{cut}$, since their time complexities have $\rho_{avg}$.
That is, as $d_{cut}$ becomes larger, their running time tends to become larger.
However, S-Approx-DPC is less sensitive to $d_{cut}$.
As $d_{cut}$ becomes larger, the number of cells in its grid $G'$ becomes smaller, i.e., the number of points that conduct a range search becomes smaller.
That is, the running time of S-Approx-DPC is influenced by $\rho_{avg}$ and $|G'|$.

\vs
\noindent
\underline{\textbf{Impact of number of threads.}}
We investigate the scalability to the number of threads.
Figure \ref{figure_thread} shows that each algorithm normally improves its running time with increase of available threads.
We see that Scan and CFSFDP-A are slow even when using 48 threads, and LSH-DDP is affected by the distribution of a given dataset.
For example, LSH-DDP scales well to the number of threads on Airline and PAMAP2 but does not on the other datasets, because it does not consider load balancing when partitioning $P$.

The limitation of Ex-DPC (dependent point computation cannot be parallelized) is observed from the result.
As we have more threads, the main overhead of Ex-DPC becomes dependent point computation, then its running time cannot be reduced.
On Sensor dataset, Ex-DPC seems to scale well, because the main cost is derived from local density computation, as shown in Table \ref{table_decomposed-time}.
On the other hand, Figure \ref{figure_thread} demonstrates that Approx-DPC and S-Approx-DPC can exploit available threads.
For example, when using 48 threads, Approx-DPC terminates clustering within 20 seconds.
Besides, in this case, Approx-DPC becomes 15.9, 24.4, 19.4, and 23.0 times faster, compared with the case of a single thread, on Airline, Household, PAMAP2, and Sensor, respectively.
S-Approx-DPC has a similar result.
(The reason why Approx-DPC and S-Approx-DPC cannot achieve 48 times faster is the hardware problem, i.e., two CPUs share a RAM and hyper-threading.)

\vs
\noindent
\underline{\textbf{Memory usage.}}
Last, we study the memory usage of the evaluated algorithms by using default setting.
Table \ref{table_memory} shows the result.
Ex-DPC consumes almost the same memory as R-tree, and our approximation algorithms need less memory than LSH-DDP and CFSFDP-A.
The memory usage of our approximation algorithms is higher than that of Ex-DPC, since they use a grid as an additional index.
In addition, $\epsilon$ is less than 1 for S-Approx-DPC, so it creates more cells than Approx-DPC.
Therefore, S-Approx-DPC needs more memory usage than Approx-DPC.

\begin{table}[!t]
\begin{center}
	\caption{Memory usage [MB]}	\label{table_memory}
	\begin{tabular}{c||c|c|c|c} \hline
                				& Airline	    & Household	    & PAMAP2        & Sensor        \\ \hline \hline
                R-tree + Scan   & 564           & 346           & 277           & 133           \\ \hline
                LSH-DDP         & 2061          & 756           & 1455          & 342           \\ \hline
                CFSFDP-A        & 59362         & 12601         & 32206         & 3900          \\ \hline
        		Ex-DPC			& \textbf{461}	& \textbf{171}	& \textbf{321}  & \textbf{93}   \\ \hline
                Approx-DPC		& 1316	        & 422	        & 790           & 201	        \\ \hline
                S-Approx-DPC	& 1410	        & 482	        & 884           & 216	        \\ \hline
	\end{tabular}
\end{center}
\end{table}

\section{Conclusion}	\label{section_conclusion}
Density-based clustering is an important technique for many data mining tasks, and Density-Peaks Clustering is one of density-based clustering techniques and has many applications.
However, its computational efficiency has not been considered much so far, although its straightforward implementation incurs $O(n^2)$ time, where $n$ is the number of input points.

In this paper, we proposed efficient Density-Peaks Clustering algorithms, Ex-DPC, Approx-DPC, and S-Approx-DPC.
As long as the average local density is sufficiently small, our algorithms are sub-quadratic.
In addition, we considered multicore-based parallel processing, which is being popularized in many systems, and presented how to parallelize our algorithms.
Because Ex-DPC cannot be fully parallelized, Approx-DPC and S-Approx-DPC are carefully designed so that they can parallelize main operations by allowing approximate results.
Our experimental results have confirmed that Ex-DPC is much faster than the state-of-the-art exact algorithm, and Approx-DPC and S-Approx-DPC are more accurate, efficient, and scalable than the state-of-the-art approximation algorithms.

\section*{Acknowledgments}
This research is partially supported by JSPS Grant-in-Aid for Scientific Research (A) Grant Number 18H04095, JST CREST Grant Number J181401085, and JST PRESTO Grant Number JPMJPR1931.

\bibliographystyle{ACM-Reference-Format}
\bibliography{sigproc}

\end{document}